%% file: article.tex
\documentclass[11pt,a4paper]{article}
\pdfoutput=1

\input{packages.tex}

\input{macros.tex}

\allowdisplaybreaks[4] 

\makeatletter
\def\@fpheader{\relax}
\makeatother

\begin{document}

\title{Kinematic cosmic dipole from a large sample of strong lenses}
  
\author[a]{Martin Millon,}
\affiliation[a]{D\'epartement de Physique Th\'eorique, Universit\'e de Gen\`eve, \\ 24 quai Ernest-Ansermet, CH-1211 Gen\`eve 4, Switzerland}
\emailAdd{martin.millon@unige.ch}

\author[b]{Charles Dalang,}
\affiliation[b]{Institut Philippe Meyer, Département de Physique,
Ecole Normale Supérieure (ENS), \\ Université PSL, 24 Rue Lhomond, F-75231, Paris, France}
\emailAdd{charles.dalang@phys.ens.fr}

\author[c]{Thomas Collett,}
\affiliation[c]{ Institute of Cosmology and Gravitation, University of Portsmouth, \\ Burnaby Road, Portsmouth PO1 3FX, UK}
\emailAdd{thomas.collett@port.ac.uk}

\author[a]{Camille Bonvin}
\emailAdd{camille.bonvin@unige.ch}

\abstract{
Measurements of the kinematic cosmic dipole continue to show an intriguing tension between the value inferred from the CMB and that obtained from high-redshift source number counts. While the measured dipole direction appears consistent, the amplitude, set by the observer's peculiar velocity $v_{o}$, remains in significant disagreement. In this paper, we propose using strong gravitational lenses with well-measured Einstein radii to estimate the kinematic cosmic dipole, through the relativistic aberration of the Einstein angle induced by the observer's motion. We show that this effect could be detected solely from measurements of the Einstein radius in wide, high-resolution imaging surveys such as Euclid. However, the precision achievable using Einstein-radius measurements alone, without redshift or lens-galaxy mass information, appears insufficient to discriminate between the CMB value of $v_{o}$ and that derived from source number counts at high statistical significance. Nevertheless, we demonstrate that including a large sample of lenses with available kinematic information, either via the Fundamental Plane relation or, ideally, through spectroscopic velocity-dispersion measurements, drastically reduces the noise and substantially improves the constraining power of this method. We show that, for a realistic sample of strong lenses detected by Euclid and complemented with spectroscopic velocity dispersion measurements from 4MOST or DESI, it is possible to discriminate between the CMB- and source-number-counts-inferred values at the $\sim 5\sigma$ level using a new, fully independent method. We further demonstrate that this technique is only weakly sensitive to strong-lensing selection effects, with selection biases and threshold effects estimated to be well below the 1\% level.}

\maketitle

\section{Introduction}

Using measurements of the dipole anisotropy in the blackbody temperature of the Cosmic Microwave Background (CMB), the \textit{Planck} satellite has accurately determined both the magnitude and direction of the observer's velocity. The observer's motion produces a temperature modulation of order $\delta T/\langle T \rangle \sim \mathcal{O}(10^{-3})$
around the angularly averaged temperature $\langle T \rangle$. This effect allows for a precise measurement of the observer's velocity, with magnitude $v_{o} = 369.82 \pm 0.11 \,\mathrm{km\,s^{-1}},$ directed towards $\hat{\bs{v}}_o = (264.021^\circ \pm 0.011^\circ, 48.253^\circ \pm 0.005^\circ)$
in Galactic coordinates \cite{Fixsen:1994, Fixsen:1996, Aghanim:2018eyx, Planck:2013kqc}. This interpretation is valid under the assumption that the intrinsic CMB dipole due to primordial density perturbations, expected to be of order $\mathcal{O}(10^{-5} - 10^{-6})$, is small in comparison. 

The observer's velocity also introduces two other signatures in the CMB: a Doppler \emph{modulation} effect, which amplifies the apparent temperature fluctuations in the velocity direction and reduces them in the opposite direction, and an \emph{aberration} effect, which changes the angular scales of the temperature fluctuations. These two effects induce correlations between the $l$ and $l \pm 1$ multipoles of the CMB. Both of these effects have been detected \cite{Planck:2013kqc, Ferreira:2020aqa, Saha:2021bay}, leading to results consistent with the velocity inferred from the dipole, and thus supporting the \emph{entirely kinematic} interpretation of the CMB dipole. However, the relatively large error bars still allow for an intrinsic component at the level of $\sim 40\%$ of the CMB dipole \cite{Schwarz:2016}. More recently, kinematic Sunyaev–Zel'dovich velocity reconstruction has been used to constrain a possible intrinsic dipole between the CMB and large-scale-structure rest frames to less than 14 $\mathrm{km\,s^{-1}}$ at 68\% confidence~\cite{Bloch2024, Krywonos2024}.

On the other hand, the observer's velocity can be measured independently using high-redshift source number counts, such as quasars, following a method originally proposed by~\cite{Ellis1984}. The observer velocity generates a dipolar modulation in the source number counts, due both to aberration and to threshold effects, i.e.\,the fact that sources will appear brighter in the velocity direction, increasing the number of sources above the flux threshold. In such analyzes, it is crucial to consider sources distributed over a sufficiently large cosmological volume in order to minimize contamination from local structures. Recent measurements based on quasar number counts from large imaging surveys point toward a discrepant value of the observer's velocity, with several analyzes reporting a tension at the level of $3$--$5.7\sigma$ \citep{Secrest:2020has, Dam:2022wwh, Secrest:2022uvx, Land-Strykowski2025} but the datasets used in these works do not always yield consistent results with, e.g., a $\sim 3\sigma$ tension between the \textsc{RACS-low} and \textsc{CatWISE} catalogs reported in \cite{Land-Strykowski2025}. Contamination by low-redshift sources affected by local clustering \cite{Oayda2024} could account for part of this discrepancy. Evolution of the source population was proposed as a possible explanation for the tension~\cite{Dalang:2021ruy}, until it was shown not to affect the number count dipole, if source properties are measured close to the flux limit \cite{vonHausegger:2024jan}. Overall, while measurements based on the \textsc{Quaia} catalog \cite{Storey-Fisher2024} seem not to support the presence of such a tension \citep{Mittal2024, Abghari2024}, a recent re-analysis of the \textsc{CatWISE} data still finds discrepant results at the $\sim 3\sigma$ level, even after accounting for possible local clustering effects \cite{Bashir2025}. Redshift tomography \cite{vonHausegger:2024fcu} and measurements of $l$ and $l+1$ correlations in source number counts  \cite{Lacasa:2024ybp, Oayda2025} were proposed as powerful complementary methods, allowing to cleanly disentangle threshold effects from the observer velocity. A review of the current status of these observations, potential systematics impacting these measurements, and the cosmological implications of the CMB dipole anomaly can be found in \cite{secrest2025, Secrest2025c}. These discrepant results from the source number-count technique highlight the need for independent methods to measure the observer's velocity.

In this work, we propose an independent measurement of the observer's peculiar velocity using strong gravitational lenses. Our method relies solely on angular measurements on the sky, providing a very robust way of extracting the observer velocity. The observable is derived purely from special relativity: a non-comoving observer perceives relativistic aberration, which induces a direction-dependent modulation of the observed Einstein radius. We demonstrate that selection effects, which modulate the number of detected Einstein rings across the sky, have a completely negligible impact on the measured dipole, below 0.1\%. As such, this observable is insensitive to flux-related and clustering systematics that may bias source-count dipole experiments. This measurement, however, requires a very large number of strong gravitational lenses to detect a small change in the Einstein ring's size for lenses aligned with the observer velocity. 

Euclid is expected to find about $\mathcal{O}(10^{5})$ strong gravitational lenses by the end of the mission, a 2 order of magnitude increase compared to the $\mathcal{O}(10^{3})$ lenses known from Stage-III imaging surveys. Already in the first Q1 data release, covering only 63 square-degrees, Euclid has revealed 497 high-quality new lens candidates \cite{EuclidWalmsley2025} thanks to the combination of computer vision methods and human visual inspection \cite{Euclid2025Rojas, EuclidLines2025, EuclidHolloway2025}. According to \cite{EuclidWalmsley2025}, this is expected to scale to approximately 7,000 new strong lenses in the first Euclid data release (DR1) and to more than $10^{5}$ lenses over the entire Euclid Wide survey. In addition, the image quality of Euclid enables very precise measurements of the Einstein angle, at the level of $\sim 1\%$. As we demonstrate in this work, the observer’s peculiar velocity can be extracted using this observable alone. If spectroscopic redshifts for the background sources and the lens galaxy, as well as velocity dispersion information for the lens are available, even for a sub-set of all systems, the precision of the measurement can be significantly improved. 

This paper is organized as follows. In Section~\ref{sec:Einsteinangle}, we derive the equations describing the aberration of the Einstein angle induced by the observer’s velocity in strongly lensed systems. In Section~\ref{sec:Lensing_only}, we introduce the Bayesian framework used to infer the observer’s velocity, present the simulated lens sample used to validate the method, and report the resulting constraints on the dipole amplitude and direction obtained using lensing information alone. In Section~\ref{sec:Lensing+kinematics}, we show how additional lens kinematic information, either from spectroscopic measurements or inferred via the Fundamental Plane (FP) relation, can be incorporated into the analysis to improve the constraints. Finally, we conclude and discuss the limitations of the method in Section~\ref{sec:discussion}. Throughout the paper, we assume the flat $\Lambda$-Cold Dark Matter model with $H_0 = 70$ \ksmpc and $\Omega_m = 0.3$ to convert redshifts into angular diameter distances. 

\section{Einstein angle aberration}
\label{sec:Einsteinangle}

\begin{figure}
    \centering
    \includegraphics[width=1.0\linewidth]{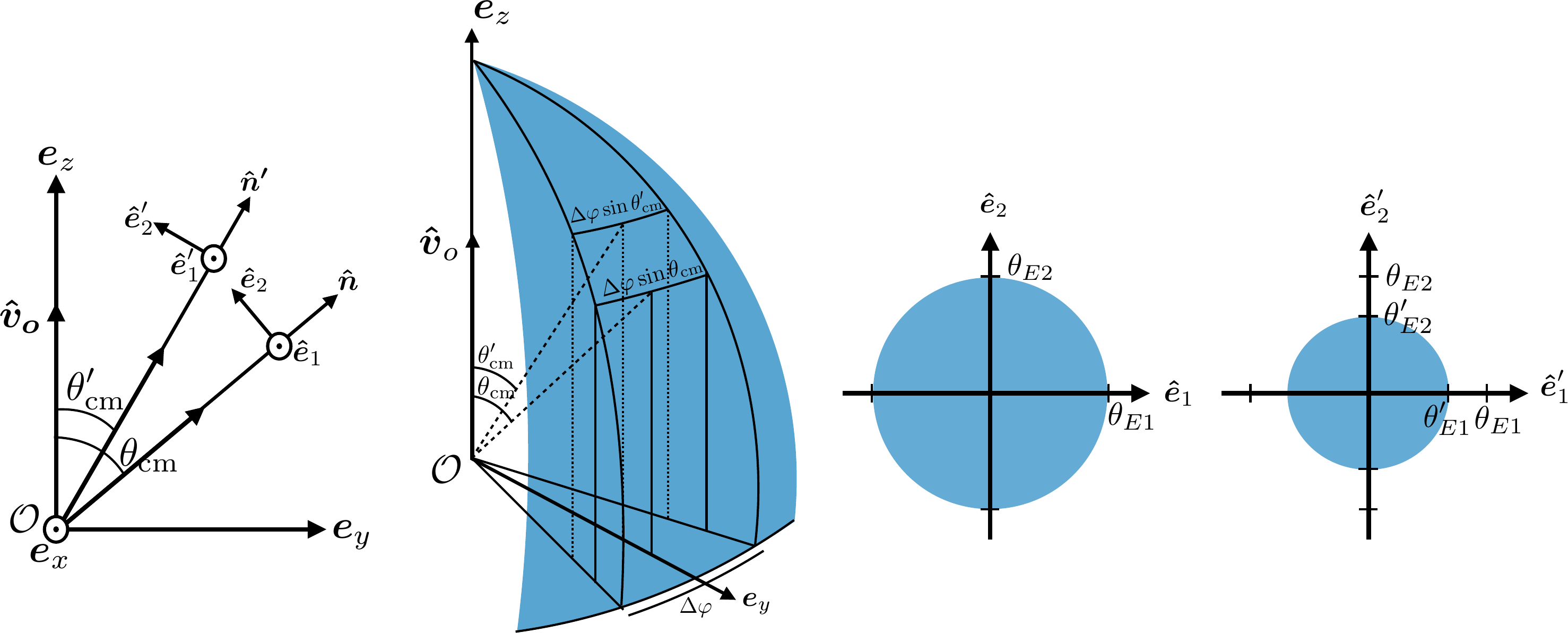}
    \caption{\textit{Left panel:} Coordinate system used in this work. We consider two observers, located at the origin $\mathcal{O}$. The first observer is stationary in this frame and measures comoving quantities, which are written without primes, e.g.\ $\theta\e{cm}$. Because of the symmetry around the $\bs{e}_z$ axis, the system is well captured by lenses which lie in the $\bs{e}_y-\bs{e}_z$ plane. The stationary observer sees a perfect circular Einstein ring with Einstein angle $\theta_E$ in the direction $\bs{\hat{n}}$. They associate a coordinate system orthogonal to $\bs{\hat{n}}$ on the sky $\{\bs{\hat{e}}_1,\bs{\hat{e}}_2\}$, depicted in the \textit{third panel}. This coordinate system is such that the $\bs{\hat{e}}_1$ vector is orthogonal both to $\bs{\hat{n}}$ and $\bs{v}_o$ (here it is aligned with $\bs{e}_x$), while $\bs{\hat{e}}_2$ lives in the $\bs{e}_y-\bs{e}_z$ plane.  While angles along $\varphi$ are not affected by the boost ($\Delta \varphi \equiv \varphi\e{max}- \varphi\e{min} = \Delta \varphi'$), the Einstein ring along $\bs{\hat{e}}_1$ is. This is due to the fact that $\theta\e{cm}'$ is affected by the boost, as may be understood from the \textit{second panel}. The second observer moves with peculiar velocity $\bs{v}_o$ aligned with $\bs{\hat{e}}_z$ and observes the Einstein ring in the $\bs{\hat{n}}'$ direction, which forms an angle $\theta'\e{cm}$ with the $\bs{e}_z$ axis. For this moving observer, the Einstein ring appears smaller, as may be seen in the \textit{fourth panel}.
    This deformation preserves the shape of the Einstein ring and is proportional to $\cos(\theta\e{cm})v_o/c$.
    }
    \label{fig:coordinate_system}
\end{figure}

The peculiar velocity of the observer, $\bs{v}_o$, leads to a change in the size of the Einstein ring due to aberration, as described in \citep{Sereno2008,Dalang2023}.
Without loss of generality, we align the $\bs{e}_z$ axis along the observer's velocity, and we place the direction of the center of mass of the lens $\bn$ in the $\bs{e}_z-\bs{e}_y$ plane, as illustrated in Figure~\ref{fig:coordinate_system}. A boosted observer will see the center of mass in the direction $\bn'\neq \bn$, which, however, still lies in the $\bs{e}_z-\bs{e}_y$ plane. The Einstein ring seen by the boosted observer lies in the plane orthogonal to $\bn'$. To describe its shape, we adopt the same coordinate system as in \cite{Dalang2023}, i.e., we define $\bs{\hat{e}}_1'$ along the $\bs{e}_x$ direction, such that angles along that axis are preserved under the boost, and $\bs{\hat{e}}_2'$ lies in the $\bs{e}_z-\bs{e}_y$ plane, orthogonally to $\bn'$. Angles along $\bs{\hat{e}}_2'$ are deformed by the boost. We denote by $\theta_{\rm cm}$ the angle between $\bs{v}_o$ and the unboosted center of mass of the lens, and by $\theta_{\rm min}$ and $\theta_{\rm max}$ the angle between $\bs{v}_o$ and the unboosted extremities of the ring. The impact of the boost on an angle $\theta$, at linear order in $v_o/c$, is given by
\begin{align}
\label{eq:thetaprime}
\theta'=\theta-\frac{v_o}{c}\sin\theta'\, .   
\end{align}
With this, we find that the boosted Einstein radius in the direction $\bs{\hat{e}}_2'$ is given by
\begin{align}
\theta'_{E2}=\theta'_{\rm max}-\theta'_{\rm cm}=\theta_{E2}\left( 1 - \frac{v_o}{c} \cos\theta'_{\rm cm} \right)=\theta'_{\rm cm}-\theta'_{\rm min}\, .
\end{align}
The fact that the boosted ring remains centered on the boosted center of mass is valid in the limit $\theta_{E2}\ll \theta_{\rm cm}$. The boosted Einstein radius in direction $\bs{\hat{e}}_1'$ is given by
\begin{align}
\label{eq:theta1prime}
\theta'_{E1}=\sin\theta'_{\rm cm}(\varphi_{\rm max}-\varphi_{\rm cm})=\sin\theta'_{\rm cm}(\varphi_{\rm cm}-\varphi_{\rm min})\, ,   
\end{align}
where $\varphi$ denotes the azimuthal angles of the center of mass and of the two extremities of the ring, which are unaffected by the boost. In the first equality, we have, as before, used that the size of the ring is much smaller than $\theta_{\rm cm}$. In our coordinate system, $\varphi_{\rm cm}=0$ and $\varphi_{\rm max}=\varphi=-\varphi_{\rm min}$. From Eq.~\eqref{eq:theta1prime}, we see that even though angles along $\bs{\hat{e}}_1'$ are unaffected by the boost, the Einstein radius along that direction is impacted due to the fact that $\theta'_{\rm cm}$ changes under the boost. This may be understood visually from the second panel of Fig.\,\ref{fig:coordinate_system}. Inserting~\eqref{eq:thetaprime} into~\eqref{eq:theta1prime} and expanding at linear order in $v_o/c$, we find
\begin{align}
\theta'_{E1}=\theta_{E1} \left( 1 - \frac{v_o}{c} \cos\theta'_{\rm cm} \right)\, .   
\end{align}
The boost therefore produces an isotropic distortion of the ring\footnote{At second order, and thus not relevant for this work, the transverse velocity between the lens and the observer can distort a perfectly circular Einstein ring into an ellipse, as shown in  \cite{Sereno2008}}. In the following, we assume that the Einstein ring is a perfect circle, arising from a spherical mass distribution, such as a Singular Isothermal Sphere (SIS), while the introduction of ellipticity in the mass profile would simply increase the noise of our measurement (see Section \ref{sec:discussion}). Since the distortion is isotropic, after the boost the ring remains a circle, but with a different radius
\begin{align}
\label{eq:Einstein_Angle}
\theta'_{E}=\theta_{E} \left( 1 - \frac{v_o}{c} \cos\theta'_{\rm cm} \right)\, .    
\end{align}
The boosted Einstein radius, $\theta'_{E}$, is the primary observable in this work. In practice, this quantity is inferred by fitting a mass model to the imaging data and computing the azimuthally averaged Einstein radius of the resulting mass profile.

To increase the statistics, we measure the mean size of all detected rings in a given solid angle $\dd\Omega'$. Since the Einstein angle depends on the redshifts of the sources $z'_s$ and of the lens $z'_l$, this introduces two integrals weighted by the number of systems observed in that solid angle $\frac{\dd N(\bn',z'_s,z'_l)}{\dd\Omega' \dd z'_s \dd z'_l}$
\begin{align}
\label{eq:average_redshift}
\langle\theta'_E (\bn') \rangle=\left(\frac{\dd N(\bn')}{\dd \Omega'} \right)^{-1}\int \dd z'_s \int \dd z'_l \frac{\dd N(\bn',z'_s,z'_l)}{\dd \Omega' \dd z'_s \dd z'_l} \theta'_E(\bn',z'_s,z'_l)\, ,   
\end{align}
with 
\begin{align}
\frac{\dd N(\bn')}{\dd \Omega'}=\int \dd z'_s \int \dd z'_l \frac{\dd N(\bn',z'_s,z'_l)}{\dd \Omega' \dd z'_s \dd z'_l} \, .   
\end{align}
To evaluate Eq.~\eqref{eq:average_redshift}, we need to express the Einstein radius as a function of the boosted redshifts $z'_s$ and $z'_l$, which introduces additional distortions. The Einstein radius indeed depends on the distances to the source and to the lens, and fixed values of $z'_s$ and $z'_l$ correspond to different distances in different directions. As shown in Appendix~\ref{app:redshift}, the Einstein radius in redshift space is given by
\begin{align}
\label{eq:theta_z_main}
\theta'_{E}(\bn',z'_s,z'_l)=& \theta_{E}(r(z'_s),r(z'_l))\left(1-\frac{v_o}{c}\cos\theta'_{\rm cm}\right)+\left((1+z_s)\frac{\partial\theta_E}{\partial z_s}+(1+z_l)\frac{\partial\theta_E}{\partial z_l} \right)\frac{v_o}{c}\cos\theta'_{\rm cm}\, ,
\end{align}
where $\theta_{E}(r(z'_s),r(z'_l))$ is the unboosted Einstein ring evaluated at the comoving distances $r(z'_s)$ and $r(z'_l)$, which do not depend on direction\footnote{Note that in the second term, we can use $z$ or $z'$ interchangeably since they multiply a term that is already linear in $v_o/c$}. From Eq.~\eqref{eq:theta_z_main}, we see that if we were to measure the Einstein ring in bins of observed redshifts, the dipolar contribution would be affected by two terms: aberration and distance perturbations. However, once we integrate over all redshifts, the second contribution cancels, and only aberration persists. To see this, we need to express the number of detected Einstein rings in terms of $z_s'$ and $z'_l$. The detailed calculation can be found in Appendix~\ref{app:redshift}. We obtain
\begin{align}
\label{eq:N_z_main}
\frac{\dd N(z'_s,z'_l,\bn')}{\dd\Omega'\dd z'_s \dd z'_l}&= \frac{\dd N(r(z'_s),r(z'_l))}{\dd \Omega \dd z_s \dd z_l}\left(1 + \frac{4 v_o}{c}\cos\theta'_{\rm cm} \right) \\
&+\left[(1+z_s)\frac{\partial}{\partial z_s}\left(\frac{\dd N}{\dd \Omega \dd z_s \dd z_l} \right) +(1+z_l)\frac{\partial}{\partial z_l}\left(\frac{\dd N}{\dd \Omega \dd z_s \dd z_l} \right)\right] \frac{v_o}{c}\cos\theta'_{\rm cm}\, . \nonumber
\end{align}
The second line in Eq.~\eqref{eq:N_z_main} combines with the $2 v_o/c$ in the first line, and with the second term in Eq.~\eqref{eq:theta_z_main} to give rise to a boundary term that exactly vanishes (see Appendix~\ref{app:redshift} for more detail). Moreover, the effect of aberration in the number of detected systems (the remaining $2 v_o/c$ in the first line of Eq.~\eqref{eq:N_z_main}) cancels in the numerator and the denominator of Eq.~\eqref{eq:average_redshift}. Due to aberration, we see more Einstein rings per solid angle in the direction of the boost, but since we normalize the Einstein radius by the number of detected Einstein rings, this effect does not impact the final result. With this, we obtain
\begin{align}
\label{eq:mean_angle_distance_z_main}
\langle\theta'_E (\bn') \rangle=\left(1 - \frac{v_o}{c}\cos\theta'_{\rm cm}\right)\langle\theta_E \rangle\, ,   
\end{align}
where $\langle\theta_E \rangle$ is the mean Einstein radius of the unboosted sample
\begin{align}
\langle\theta_E \rangle=\int \dd z'_s \int \dd z'_l f(z'_s,z'_l) \theta_E(z'_s,z'_l)\, ,    
\end{align}
with $f(z'_s,z'_l)$ the redshift distribution of systems
\begin{align}
\label{eq:f_z}
f(z'_s,z'_l)=\frac{\frac{\dd N(z'_s,z'_l)}{\dd \Omega \dd z_s \dd z_l}}{\int \dd z'_s \int \dd z'_l\frac{\dd N(z'_s,z'_l)}{\dd \Omega \dd z_s \dd z_l}}\, .    
\end{align}

Eq.~\eqref{eq:mean_angle_distance_z_main} allows us to measure the observer velocity $v_o$ by fitting a monopole term, proportional to the mean of the lens sample $\langle\theta_E \rangle$, and a dipole term proportional to $v_o$. This relies on the underlying assumptions that there are no selection effects, i.e.\ that all systems are detected. In Section~\ref{sec:discussion}, we will relax this assumption, introducing a selection in flux and in size. We will show that this introduces additional contributions to the dipole, which are, however, much smaller than the effect of aberration. More precisely, the selection in size introduces a contamination of the order of 0.1\%, while the selection in flux is three orders of magnitude smaller. This shows that Einstein rings provide a very robust observable to measure the observer velocity, without being contaminated by selection effects that are typically difficult to characterize.

Note that Eq.~\eqref{eq:mean_angle_distance_z_main} has been derived in redshift space, since the source and lens redshifts can be measured. However, since we integrate over all systems, we could have instead chosen to integrate over comoving distances $r_s$ and $r_l$, without changing the result. In this case, the Einstein radius is simply given by Eq.~\eqref{eq:Einstein_Angle}, and the number of systems is only affected by aberration $\dd \Omega/\dd \Omega'=1+2\frac{v_o}{c}\cos\theta'_{\rm cm}$. The final result is then the same as Eq.~\eqref{eq:mean_angle_distance_z_main}, with redshifts replaced by comoving distances.

\section{Measurement of the observer velocity from lensing alone}
\label{sec:Lensing_only}
\subsection{Likelihood definition}

We consider a catalog of $N$ azimuthally averaged Einstein radii measurements $\theta'_{E_n}$ with associated galactic coordinates $(l_n,b_n)$, forming an angle $\theta'_{{\rm cm}_{,n}}$ with respect to the dipole direction ($l_o$, $b_o$). We define $\langle\theta_E\rangle$ as the mean Einstein radius of the sample, $\sigma_{\theta_E}$ as the standard deviation of the distribution, and $\delta \theta'_{E_n}$ as the associated measurement error for each lens. One can fit a monopole and a dipole by maximizing the likelihood: 
\begin{equation}
\label{eq:probe1_logl}
\ln \mathcal{L}_1
= -\frac{1}{2}\sum_{n=1}^{N}
\frac{\left[\theta'_{E_n}
- \langle\theta_E\rangle\left(1 
- \frac{v_o}{c}\cos (\theta'_{{\rm cm}_{,n}})
\right)\right]^2}
{\delta \theta^{'2}_{E_n} + \sigma_{\theta_E}^2}
- \frac{1}{2}\sum_{n=1}^{N}
\ln\!\big[2\pi\big(\delta \theta^{'2}_{E_n} + \sigma_{\theta_E}^2\big)\big]\, ,
\end{equation}
where $\langle\theta_E\rangle$, $\sigma_{\theta_E}$, $l_o$, $b_o$, and $v_o$ are free parameters.

Here, we fit a displacement of the mean of the Einstein–radius distribution as a function of position on the sky. This probe has the advantage that it does not require any spectroscopic information about the lenses, not even the lens or source redshifts. The measurement relies solely on angle measurements in the plane of the sky. However, the intrinsic scatter of the distribution, $\sigma_{\theta_E}$, constitutes a significant source of noise, driven primarily by the broad distribution of lens galaxy masses and the wide range of lens and source redshifts. The noise introduced by the unknown lens mass is analogous to the unknown intrinsic shape of the galaxy, introducing shape noise in weak lensing analysis. However, in contrast to weak lensing, the mass of the lensing galaxy can be estimated independently from other techniques. As we demonstrate in Section \ref{sec:Lensing+kinematics}, incorporating spectroscopic redshift information for both the lens and the source, together with independent mass estimates (e.g., from velocity–dispersion measurements or from the FP relation; \cite{Dressler1987, Djorgovski1987}), can substantially reduce this noise.

\subsection{Lens population model}
\label{subsec:lenspop}
We use the {\sc LensPop} \citep{LensPop,Collett2015} lens simulation package to
generate a realistic population of strong lenses expected from Euclid.
{\sc LensPop} draws deflector galaxies from the observed distributions of
velocity dispersions and ellipticities measured in the Sloan Digital Sky
Survey~\citep{Park2007}. Deflectors are assumed to be isothermal and
are randomly distributed in comoving volume. Sources are placed behind the
deflectors, with source redshifts and luminosities drawn from the catalog of
\cite{Connolly2010}, which uses semi-analytic models to populate galaxies onto
the dark matter halos of the Millennium Simulation
\citep{DeLucia2006}.

This procedure results in a population of lenses representative of those
present in the Universe; however, for our purposes, we require the subset of
lenses that would be {\it discovered} by Euclid. To this end, {\sc LensPop}
simulates observations of this population with Euclid-like data quality and
applies a simplified discovery selection function. As described in
\cite{Collett2015}, this selection function can be summarized by requirements
on the signal-to-noise ratio of the source, a minimum Einstein radius, a total
source magnification greater than 3, and an arc length sufficient for the
system to be resolved as an extended object. 

\begin{figure} [h!]
    \centering
    \includegraphics[width=\linewidth]{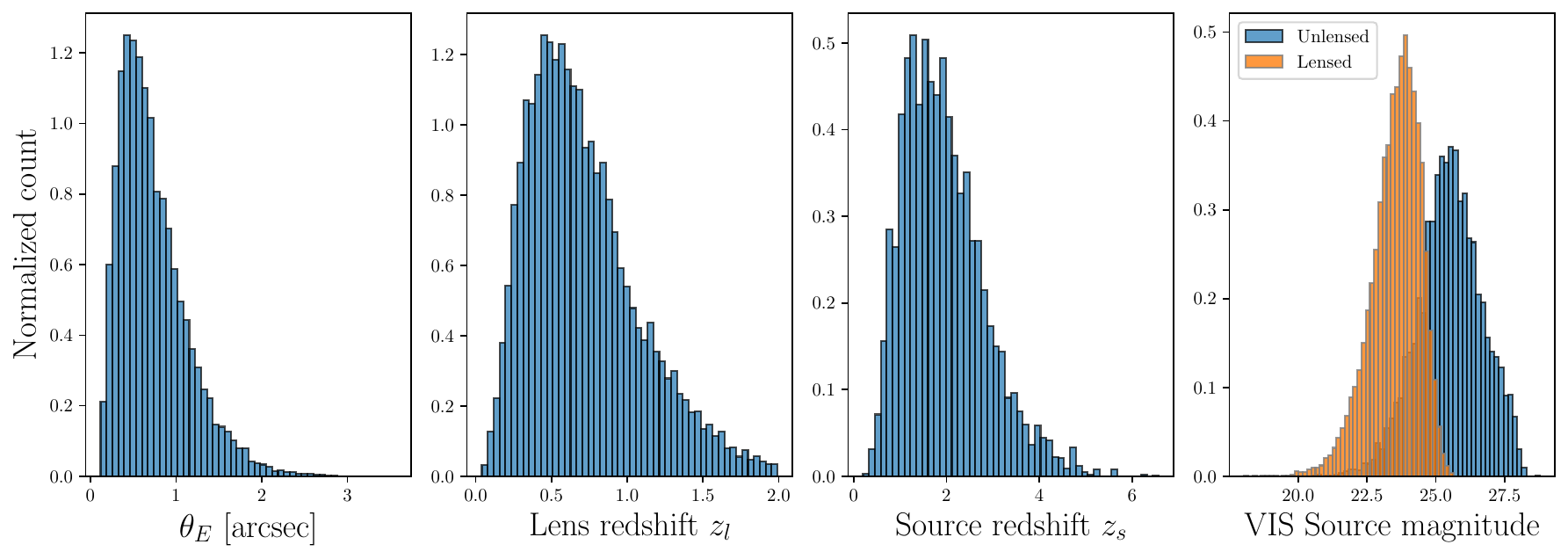}
    \caption{Properties of the simulated Euclid lens sample considered in this work.}
    \label{fig:lens_sample}
\end{figure}

The properties of the resulting lens sample are shown in
Figure~\ref{fig:lens_sample}. While the simplifying assumptions of {\sc LensPop}
cannot perfectly reproduce the true Euclid selection function, they have been
shown to accurately predict both the number of galaxy-galaxy lenses
discoverable by Euclid and their Einstein radius distribution
\citep{AcevedoBarroso2025,EuclidWalmsley2025}. These are precisely the
quantities that must be modeled reliably for our cosmic dipole inference
forecast to be robust.

From this sample, we randomly assign galactic coordinates $(l_n, b_n)$ to each lens in the catalog within a footprint that mimics the Euclid Wide Survey footprint, requiring $|b_n| > 30$\degree. We simulate the observed $\theta'_{E_n}$ from the true $\theta_{E_n}$ using Eq.~\eqref{eq:Einstein_Angle} and adding Gaussian noise with standard deviation $\delta \theta'_{E_n} / \theta'_{E_n}$ = 0.02. Euclid could measure the Einstein radius with a precision of about 1\%, however, we adopt a relative 2\% uncertainty to account for additional error contributions from line-of-sight large-scale structures.

\subsection{Lensing only results}
\begin{figure}[h!]
    \centering
    \includegraphics[width=0.8\linewidth]{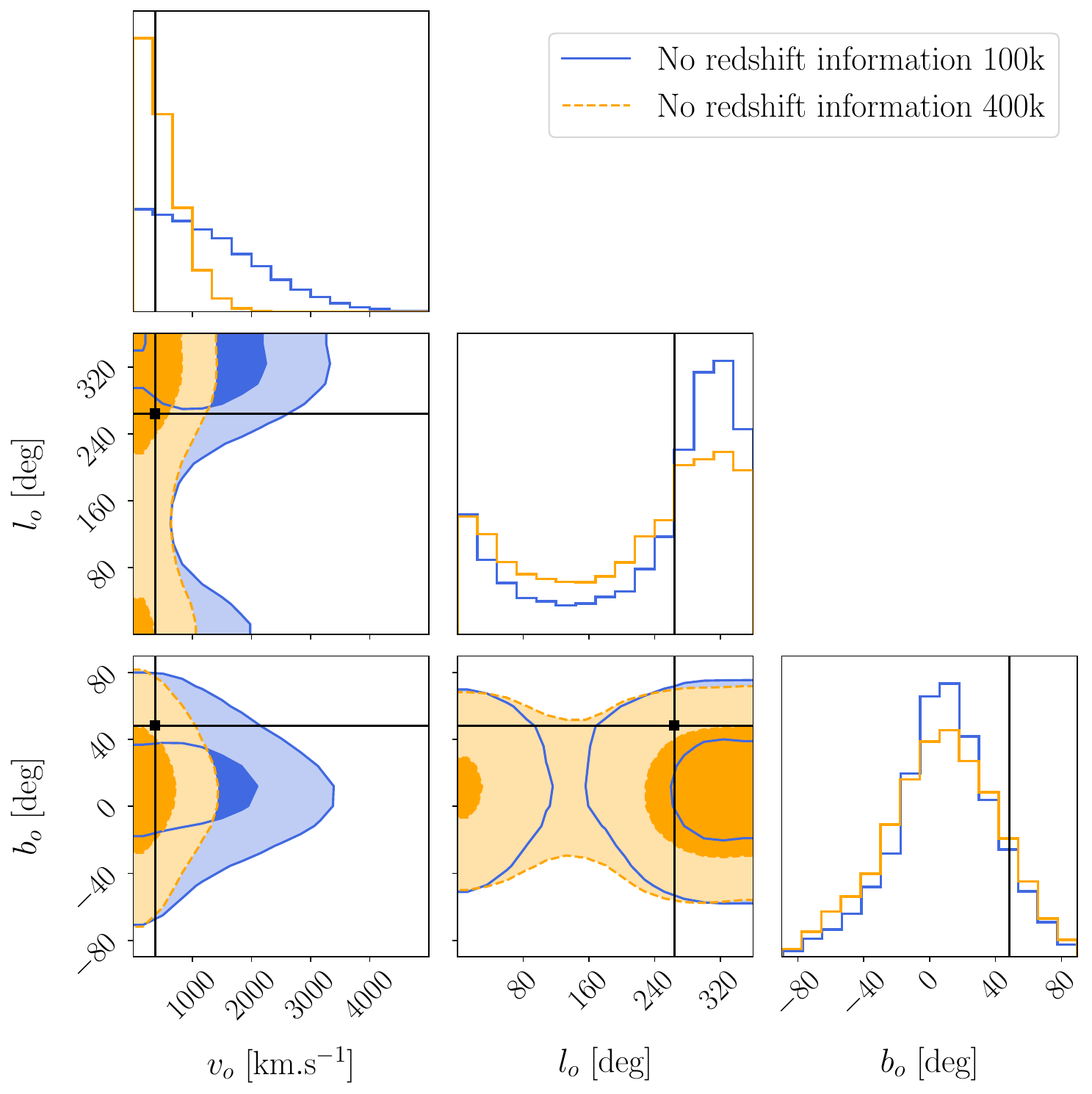}
    \caption{
    Posterior probability distributions of the kinematic dipole amplitude and direction inferred from samples of $100\,000$ (blue contours) and $400\,000$ (orange contours) strong lenses with Euclid-quality imaging. The black lines indicate the true value of $v_{o} = 369.82$ \ks, pointing toward Galactic coordinates $(l_o,b_o) = (264.021^\circ,\; 48.253^\circ)$. The contours enclose 39.3\% and 86.5\% of the posterior probability.
}
    \label{fig:lensing_only_posterior}
\end{figure}

From the simulated lens catalog described in Section~\ref{subsec:lenspop}, we randomly select samples of $100\,000$ (the ``100k" sample) and $400\,000$ lenses (``400k" sample), from which we evaluate the lensing-only likelihood (Eq.~\ref{eq:probe1_logl}). We then infer the posterior probability distributions of the parameters $v_o$, $l_o$, and $b_o$ using the \textsc{Nautilus} sampler \cite{Nautilus}. The $1\sigma$ and $2\sigma$ two-dimensional posterior probability contours for these parameters are shown in Fig.~\ref{fig:lensing_only_posterior}. The remaining free parameters, $\langle\theta_E\rangle$ and $\sigma_{\theta_E}$, are marginalized over and are not shown in this figure. Table~\ref{tab:lensing_only_results} summarizes the marginalized posterior constraints obtained both with and without fixing the dipole direction. We 
repeat the analysis for two different true values of the observer’s peculiar velocity: $v_{o,\rm true} = 369.82$\ks, corresponding to the CMB-inferred value, and $v_{o,\rm true} = 1109.46$ \ks, corresponding to three times the CMB value and consistent with recent results from source number counts \citep[e.g.][]{Secrest:2022uvx}.

While the dipole is detected in both the ``100k" and ``400k" samples for the highest peculiar velocity, $v_{o,\rm true} = 1109.46$\ks, the lowest CMB-inferred value does not produce a sufficiently strong signal to be detected in the ``100k" sample. Fixing the dipole direction leads to a moderate tightening of the constraints on the dipole amplitude, improving the 68\% credible interval width on $v_{o}$ by approximately 15 to 30\%, depending on the sample size and fiducial velocity.

We therefore find that, when relying on lensing information alone, the Euclid lens sample is unlikely to provide strong discriminatory power between the CMB-inferred dipole amplitude and the larger values suggested by source number count analyzes. The resulting constraints are not sufficiently precise to exclude the CMB-inferred value beyond the $\sim 1.0\sigma$ level, even when fixing the dipole direction. For the ``400k" sample, the discriminating power reaches $\sim 2\sigma$. We explore the possibility of including the stellar kinematic information to improve the precision of this measurement in the next section. 

\begin{table}[h!]
\centering
\resizebox{\linewidth}{!}{
\renewcommand{\arraystretch}{1.1}
\begin{tabular}{l|ccc|ccc}

& \multicolumn{3}{c}{$v_{o,\rm true} = 369.82$ \ks} &  \multicolumn{3}{|c}{$v_{o,\rm true} = 1109.46$ \ks} \\ \hline
Lens sample & $v_{o}$ [km s$^{-1}$] & $l$ [deg] & $b$ [deg] & $v_{o}$ [km s$^{-1}$] & $l$ [deg] & $b$ [deg] \\ \hline

Lensing only 100k & $1103^{+1109}_{-775}$ & $284.9^{+47.1}_{-212.0}$ & $9.6^{+29.1}_{-28.5}$ & $1695^{+1180}_{-1057}$ & $292.6^{+35.9}_{-97.9}$ & $22.6^{+23.6}_{-21.5}$ \\
Lensing only 400k & $393^{+444}_{-273}$ & $247.6^{+77.7}_{-190.3}$ & $8.4^{+33.1}_{-35.2}$ & $974^{+530}_{-500}$ & $282.5^{+38.2}_{-63.4}$ & $33.5^{+21.7}_{-20.5}$ \\
Lensing only 100k - Fixed direction & $943^{+737}_{-602}$ & -- & -- & $1555^{+767}_{-749}$ & -- & -- \\
Lensing only 400k - Fixed direction & $394^{+336}_{-260}$ & -- & -- & $1002^{+401}_{-386}$ & -- & -- \\

\end{tabular}}
\caption{Posterior constraints on the kinematic dipole amplitude and direction using only lensing information. The second and third columns correspond to fiducial values of $v_{o}$ equal to the CMB-inferred value and three times the CMB value, respectively. The fourth and fifth rows of the table assume a fixed dipole direction toward $(l_o,b_o) = (264.021^\circ,\; 48.253^\circ)$. Reported values correspond to the 16$^{\rm th}$, 50$^{\rm th}$, and 84$^{\rm th}$ percentiles.}

\label{tab:lensing_only_results}
\end{table}

\section{Combining lensing and Stellar Kinematics}
\label{sec:Lensing+kinematics}

The dominant source of uncertainty in the measurement of the observer velocity arises from the width of the Einstein radius distribution, $\sigma_{\theta_E}$. This uncertainty can be reduced by incorporating additional information that constrains the unboosted Einstein radius of each ring. We explore two sets of complementary datasets here.

\subsection{Spectroscopic velocity dispersion}

First, the stellar velocity dispersion of the lens in combination with the source and lens redshifts,  provides information about its mass, which directly informs the Einstein radius. More precisely, for a SIS, the central stellar velocity dispersion of the lens galaxy $\sigma_v$, relates to the Einstein radius through: 
\be
\label{eq:SIS}
\sigma_v = c \, \sqrt{ 
    \frac{\theta_{E}}{4 \pi} \, \frac{d_s}{d_{ls}}\,,
}
\ee
where $d_s$ and $d_{ls}$ are the angular diameter distances between the observer and the source and between the lens and the source, respectively. The dispersion $\sigma_v$ is measurable from long-slit or fiber spectroscopic observations of the lens galaxy. It is unaffected by the observer's peculiar velocity because it is a relative measurement of wavelength dispersion. This additional observable provides an estimate of the mass of each lensing galaxy, hence largely reducing the noise term in Eq.~\eqref{eq:probe1_logl}. The fractional measurement error on $\sigma_v$ translates to $\theta_{E}$ according to: 
\be 
    \frac{\delta \sigma_v}{\sigma_v} = \frac{1}{2} \frac{\delta \theta_{E}}{\theta_{E}}\,.
\ee
However, the angular diameter distances $d_{ls}$ and $d_s$ in Eq.~\eqref{eq:SIS} have to be inferred from the observed redshifts $z_l'$, $z_s'$, and a cosmology, which are biased by the peculiar velocities of the observer, of the lens, and of the source. Using the same notation as in~\cite{Dalang2023}, the observed redshifts $z_l'$, $z_s'$ relate to the cosmological (or background) redshift $z_l$, $z_s$ in the following way 
\begin{align}
(1+z_l) & = (1+z_l')\left(1+Z_L \frac{v_o}{c} \right)\,, \label{eq:zl}\\
(1+z_s) & = (1+z_s')\left(1+Z_S \frac{v_o}{c} \right) \label{eq:zs}\,,
\end{align}
with
\begin{align}
Z_L & =\frac{\bs{\hat{n}'} \cdot (\bs{v}_o - \bs{v}_l)}{v_o} \,, \label{eq:Zl}\\
Z_S & = \frac{\bs{\hat{n}'} \cdot (\bs{v}_o - \bs{v}_s)}{v_o}\,.\label{eq:Zs}
\end{align}
Propagating these corrections to the background angular diameter distances, one finds, to linear order in $v_o/c$ \cite{Dalang2023}
\begin{align}
d_l &  = d[z_l'] + D_L \frac{v_o}{c}\,,\label{eq:dl_1}\\
d_{ls}&  = d[z_l',z_s'] + D_{LS} \frac{v_o}{c}\,,\label{eq:dls_1}
\end{align}
with the distance corrections 
\begin{align}
D_L & =  \frac{\bs{\hat{n}'}\cdot(\bs{v}_o-\bs{v}_l)}{v_o} \left( \frac{c}{H[z_l']}  - d[z_l']\right)\,, \label{eq:DL} \\
D_{LS} & =\frac{\bs{\hat{n}'}\cdot(\bs{v}_o-\bs{v}_s)}{v_o} \left(  \frac{c}{H[z_s']}  -d[z_l',z_s'] \right) - \frac{c }{H[z_l']} \frac{1+z_l'}{1+z_s'} \frac{\bs{\hat{n}'}\cdot(\bs{v}_o-\bs{v}_l)}{v_o} \,. \label{eq:DLS}
\end{align}
Here $H[z]= H_0 E(z) \equiv H_0 \sqrt{\Omega_{m}(1+z)^3 + (1-\Omega_{m})}$ and $\bs{v}_l$, $\bs{v}_s$ are the peculiar velocities of the lens and the source galaxy, respectively. Note that Eqs.~\eqref{eq:dl_1} to~\eqref{eq:DLS} do not represent the observed angular diameter distances of a moving source and lens measured by a boosted observer, since those would also be affected by changes in the area distance, see e.g.~\cite{Cusin:2024git}. The aim of these expressions is to allow us to find the background angular diameter distances $d_l$ and $d_{ls}$ (that enter in Eq.~\eqref{eq:SIS}) from the measured redshifts $z'_l$ and $z'_s$, consistently accounting for the impact of velocities.

The relationship between the observable quantities $z_l'$, $z_s'$, $\sigma_v$ and the predicted Einstein radius $\theta'_{E, \rm spec}$ is then given by 
\begin{align}
\label{eq:theta_spec}
\theta'_{E, \rm spec} = \frac{4 \pi \sigma_v^2}{c^2} \frac{d[z_l',z_s']}{d[z_s']} \left( 1 -\frac{v_o}{c}\cos\theta'_{\rm cm} + \frac{D_{LS}}{d[z_l',z_s']} \frac{v_o}{c} - \frac{D_S}{d[z_s']} \frac{v_o}{c} \right),
\end{align}
which can be compared to the observed $\theta_E'$. Contrary to the previous case, our aim here is not to compare each observed ring to the theoretical prediction for the mean $\langle \theta'_E (\bn')\rangle$ averaged over all source and lens redshifts in a given solid angle. Since we do know the redshifts of the sources and of the lenses, we can improve our analysis by comparing each ring with the theoretical prediction at the correct redshifts, given by Eq.~\eqref{eq:theta_spec}. In this case, the likelihood becomes: 
\begin{equation}
\label{eq:probe2_logl}
    \ln \mathcal{L}_2 = -\frac{1}{2}\sum_{n=1}^{N} \frac{\left(\theta'_{E_n} -  \theta'_{E_{n}, \rm spec}  - b_{\rm VD} \right)^2}{\delta \theta^{'2}_{E_n} + \delta \theta^{'2}_{E_{n}, \rm spec} + \sigma_{\rm sys,VD}^2}
     - \frac{1}{2}\sum_{n=1}^{N} \ln\!\big[2\pi\big(\delta \theta^{'2}_{E_n} + \delta \theta^{'2}_{E_{n}, \rm spec} + \sigma_{\rm sys,VD}^2 \big)\big]\,,  
\end{equation}
where we include a bias and a systematic error term, $b_{\rm VD}$ and $\sigma_{\rm sys,VD}$, which are fitted as free parameters. When simulating our lens sample, all peculiar velocity terms of the lens, the source, and the observer (Eqs.~\eqref{eq:DL} and \eqref{eq:DLS}) are included in the simulation. We assume that both the source and lens galaxies have peculiar velocities that are normally distributed, with random orientations and a standard deviation of $300$\ks. However, at inference, only the terms involving $\bs{v}_o$ can be corrected for (since $\bs{v}_o$ also enters into the dipole), while the terms involving $\bs{v}_l$ and $\bs{v}_s$ are not observable. This source of uncertainty, due to the random peculiar velocities of the lens and the source, is not expected to produce a dipole and is therefore absorbed into the bias and systematic error terms fitted to the data, which also absorb the redshift measurement error, assumed to be $\delta z_{\rm spec}/z_{\rm spec} =0.001$. This error term is small in comparison to the velocity dispersion measurement error. 

The uncertainties on the Einstein radius derived from the velocity dispersion measurements, $ \theta'_{E_{n}, \rm spec}$, are propagated from the uncertainties on $\sigma_v$, which we take to be $7\%$. This translates into a $14\%$ relative uncertainty on $ \theta'_{E_{n}, \rm spec}$. The $7\%$ uncertainty on $\sigma_v$ itself combines a $5\%$ contribution from measurement errors, which is conservative compared to the most recent measurement \citep{Knabel2025, TDCOSMO2025} and a $5\%$ contribution from modeling uncertainties, the latter reflecting possible departures of the mass profile from perfect sphericity or isothermality (see Section \ref{sec:discussion} for a discussion of this issue). The modeling uncertainties could, in theory, be reduced by a more complex axisymmetric Jeans modeling of the lenses instead of the simple SIS assumption, which is sufficient for our forecast.

\subsection{Fundamental Plane}
Although some spectroscopic follow-up observations of strong lenses are already planned within 4MOST \citep{4SLSLS}, the majority of the Euclid lens sample is unlikely to have spectroscopic velocity-dispersion measurements. Such measurements from ground-based telescopes are expected to be feasible only for the brightest lens galaxies in the sample. However, a larger fraction of the system will have spectroscopic source and lens redshift, or at least, reliable photometric redshifts. An estimation of the central velocity dispersion of the galaxy can be obtained through the Fundamental Plane (FP) relation, which relates the effective radius of the lens $R_e$, the central velocity dispersion $\sigma_{\mathrm{FP}}$ and the mean surface brightness within the effective radius $I_e$ as
\begin{align}
\log_{10} R_{e} &= a\log_{10} \sigma_{\mathrm{FP}} + b\log_{10} I_{e} + c\,.
\label{eq:FP_main}
\end{align}
Only the angular size of the effective radius can be directly measured from the imaging data, but $R_e$ can be straightforwardly derived from this measurement and from the (photometric) redshift of the lens, assuming a cosmological model. We assume $a=1.177$, $b=-0.793$ and $c=-0.206$, obtained from the latest DESI measurement \citep{Said2025}. Measurement errors on these parameters are ignored, as they lead to errors much smaller than the intrinsic scatter of the FP. 
While $I_e$ is invariant under a boost (as shown in the App.\,\ref{ap:B}), $R_e$ is not. It can be computed from the observed subtended angle $\alpha'$ and lens redshift $z_l'$, which are affected by the boost, as
\begin{align}
R_e = \alpha d_l = \alpha' d[z_l'] \l (1+ \frac{v_o}{c}\cos(\theta\e{cm}') + \frac{D_L}{d[z_l']} \frac{v_o}{c}\r)\,.
\end{align}
As a result, the velocity dispersion $\sigma\e{FP}$ determined from the FP is also affected by the peculiar velocities, as discussed in detail in App.\,\ref{ap:B}. On average, the observed Einstein angle $\theta'\e{E} $ can be expressed as
\begin{align}
\theta'_{E, \rm FP}= \frac{4 \pi \sigma_{\rm FP}'^2 }{c^2} \frac{d[z_l',z_s']}{d[z_s']} \l(1 - \frac{v_o}{c}\cos(\theta'\e{cm})+ \frac{D_{LS}}{d[z_l',z_s']} \frac{v_o}{c} - \frac{D_{S}}{d[z_s']} \frac{v_o}{c}  + 2 \Sigma_v \frac{v_o}{c} \r) \,,\label{eq:FP_Einstein_angle}
\end{align}
where $\sigma_{\rm FP}'$ and $\Sigma_v$ are defined as
\begin{align}
\sigma_{\rm FP}' & = \l( 10^{-c} \alpha' d[z_l'] (I_e')^{-b}\r)^{1/a} \,,\\
\Sigma_v & = \l(\frac{1}{a} \cos(\theta\e{cm}') +\frac{D_L}{a d[z_l']} \r) \,.
\end{align}
Eq.\,\eqref{eq:FP_Einstein_angle} can be used to infer $v_o$ from the observables $\theta'\e{E}$, $z_l'$, $z_s'$, $\alpha'$, $I_e'$, $\theta\e{cm}'$ and the parameters $a,b,c$. In practice, it is, however, more convenient to define the likelihood in terms of the velocity dispersion rather than the Einstein radius, i.e., to compare $\sigma_{\rm FP}'$ with the velocity dispersion $\sigma'_{\rm SIS}$ inferred from $\theta'_{E}$ using Eq.~\eqref{eq:SIS}. $\sigma_{\rm FP}'$ indeed has a large scatter, which is squared in Eq.~\eqref{eq:FP_Einstein_angle}, breaking down the assumption of a Gaussian likelihood. The likelihood function is hence defined as:

\begin{align}
\label{eq:logl_probe3}
\ln \mathcal{L}_3
= -\frac{1}{2} \sum_{n=1}^{N} 
\frac{\Big(\sigma'_{{\rm SIS},n} - \sigma'_{\mathrm{FP},n} - b_{\rm FP} \Big)^2}
{\delta \sigma_{\mathrm{SIS}, n}^2 + \delta \sigma_{\mathrm{FP},n}^2  + \sigma_{\rm sys, FP}^2}  - \frac{1}{2} \sum_{n=1}^{N} 
\ln \Big[ 2 \pi \left( \delta \sigma_{\mathrm{SIS}, n}^2 + \delta \sigma_{\mathrm{FP},n}^2  + \sigma_{\rm sys, FP}^2 \right) \Big].
\end{align}
Details on how the uncertainties $\delta \sigma_{\mathrm{FP}}$ are calculated can be found in App.\,\ref{ap:A}. Similarly to Eq.~\eqref{eq:probe2_logl}, we introduced a bias and systematic error term to account for model mismatch, peculiar velocity effects, and redshift measurement errors when converting Einstein angles to velocity dispersions. We treat separately the likelihood contributions from lenses with only photometric redshift estimates and those with spectroscopic redshift measurements for both source and lens, since these two subsamples require different bias and systematic error corrections due to the differing precision of their redshift determinations. 

All variables and priors used in the simulation and in the inference are listed in Table \ref{tab:variable_simu}. We note that the priors on the dipole direction are taken to be uniform on the sphere.

\begin{table}[]
    \centering
    \resizebox{\linewidth}{!}{
    \begin{tabular}{c|cl}
       Simulation & Value & Description  \\ \hline 
       $\delta \theta'_{E} / \theta_E'$ & 2\% & Measurement error on $\theta'_{E}$ from the Euclid images \\
        $\delta \sigma_v / \sigma_v$ & 7\% & Relative error on the spectroscopic measurement of $\sigma_v$, Ref. \cite{Knabel2025, TDCOSMO2025} \\
        $\delta z_{\rm spectro}/(1 + z_{\rm spectro})$ & 0.001 & Spectroscopic redshift measurement error, Ref. \cite{EuclidSciRD2010}\\
        $\delta z_{\rm photo}/(1 + z_{\rm photo}$) & 0.03 & Photometric redshift measurement error, Ref. \cite{EuclidSciRD2010} \\
        $v_{\rm pec}$ & 300 \ks & Standard deviation of the peculiar velocity galaxies \\ 
        $a$ &   1.177 & FP coefficient (Eq. \ref{eq:FP_main}), Ref. \cite{Said2025} \\
        $b$ &   -0.793 & FP coefficient (Eq. \ref{eq:FP_main}), Ref. \cite{Said2025} \\
        $c$ &   -0.206 & FP coefficient (Eq. \ref{eq:FP_main}), Ref. \cite{Said2025} \\
        $\sigma_1$ &   0.059 & Scatter perpendicular to the FP (Eq. \ref{eq:Sigma_rsi}), Ref. \cite{Said2025}\\
        $\sigma_2$ &   0.392 & Scatter along the FP (long axis, Eq. \ref{eq:Sigma_rsi}), Ref. \cite{Said2025} \\
        $\sigma_3$ &   0.256 & Scatter along the FP (short axis, Eq. \ref{eq:Sigma_rsi}), Ref. \cite{Said2025}\\
        $\bar{s}$ &   2.102 & Mean $s$ of the FP (Eq. \ref{eq:FP_mean}), Ref. \cite{Said2025} \\
        $\bar{i}$ &   2.653 & Mean $i$ of the FP (Eq. \ref{eq:FP_mean}), Ref. \cite{Said2025} \\
        $\bar{r}$ &   0.165 & Mean $r$ of the FP (Eq. \ref{eq:FP_mean}), Ref. \cite{Said2025} \\
        $\epsilon_r$ & 0.041 dex & Measurement error on $r$ (Eq. \ref{eq:FP_error_matrix}), \cite{Magoulas2012} \\
        $\epsilon_i$ & 0.060 dex & Measurement error on $i$ (Eq. \ref{eq:FP_error_matrix}), \cite{Magoulas2012} \\ \hline
       Inference & Prior &  \\ \hline
       $|\bm{v}_o|$ & $\mathcal{U} \sim (0, 5000)$ \ks & Velocity of the observer \\
        $l_o$ & $\mathcal{U} \sim (0, 360)$ deg & Direction of the observer’s velocity in galactic coordinates \\
        $b_o$ & $\sin(b_o) \sim \mathcal{U} (-1, 1)$ & Direction of the observer’s velocity in galactic coordinates (isotropic on the sphere)\\
        $\bar{\theta}_E$ & $\mathcal{U} \sim (0, 2)$\arcsec & Mean Einstein radius of the sample \\
        $\sigma_{\bar{\theta}_E}$ & $\mathcal{U} \sim (0, 1)$\arcsec  & Standard deviation of the Einstein radius distribution \\
        $b_{\rm VD}$ & $\mathcal{U} \sim (-1, 1)$\arcsec & Bias in spectroscopic determination of $\langle \theta_E'\rangle_{\rm spec}$ \\
        $b_{\rm FP-spectro}$ & $\mathcal{U} \sim (-300, 300)$\ks & Bias in FP determination of $\sigma_{\rm FP}$ (spectroscopic redshifts)\\
        $b_{\rm FP-photo}$ & $\mathcal{U} \sim (-300, 300)$\ks & Bias in FP determination of $\sigma_{\rm FP}$ (photometric redshifts) \\
        $\sigma_{\rm sys,\ VD}$ & $\mathcal{U} \sim (0, 1)$\arcsec & Systematic error in  $\langle \theta_E'\rangle_{\rm spec}$ \\
        $\sigma_{\rm sys,\ FP-spectro}$& $\mathcal{U} \sim (0, 300)$ \ks & Systematic error $\sigma_{\rm FP}$ (spectroscopic redshifts) \\
        $\sigma_{\rm sys,\ FP-photo}$ & $\mathcal{U} \sim (0, 300)$ \ks & Systematic error $\sigma_{\rm FP}$ (photometric redshifts) \\

    \end{tabular}}
    \caption{List of variables used in the simulation of the lens sample and observables, together with the priors adopted for the inference.}
    \label{tab:variable_simu}
\end{table}

\subsection{Euclid Forecast}

\begin{figure}
    \centering
    \includegraphics[width=0.8\linewidth]{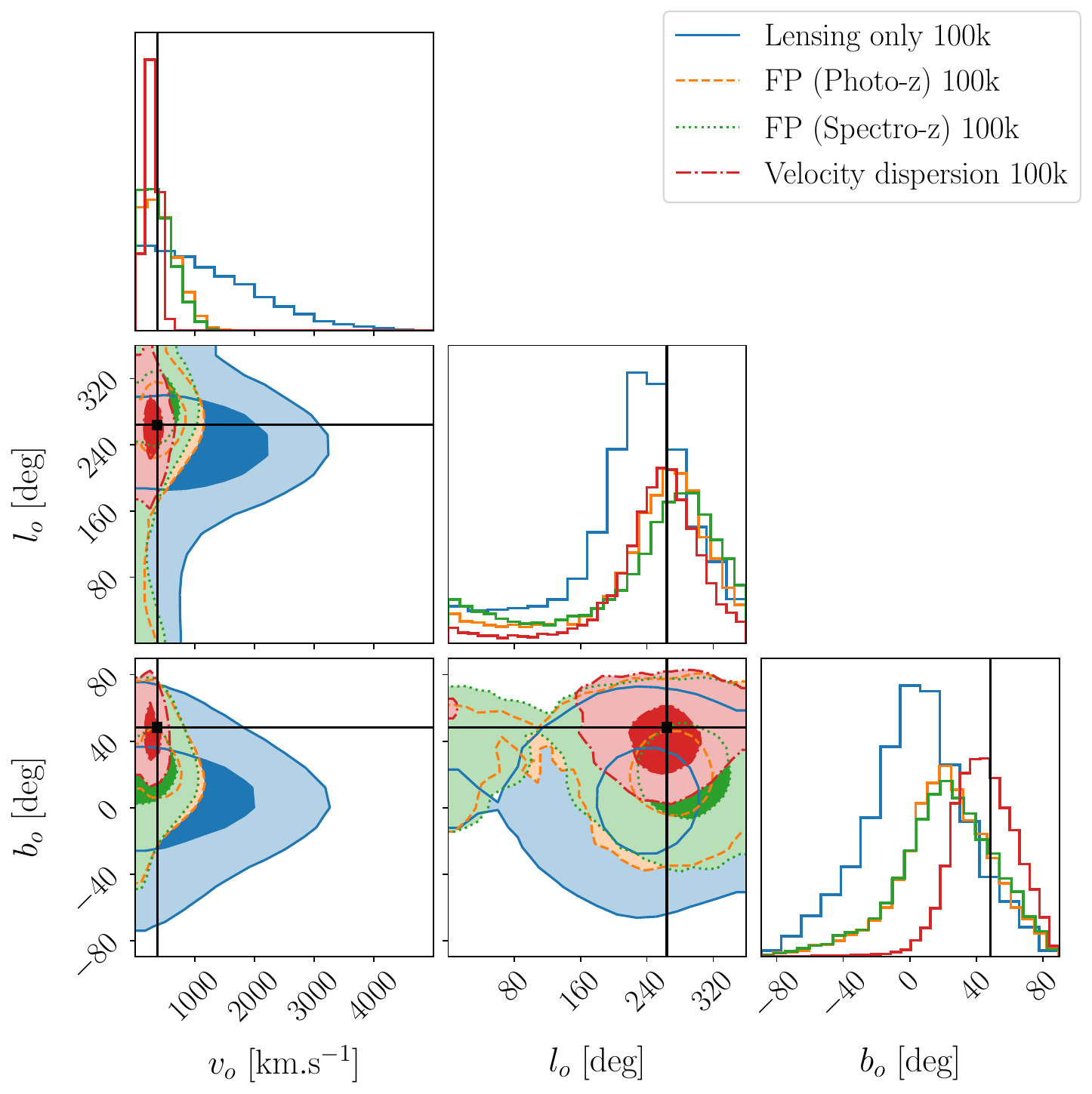}
    \caption{Posterior probability distributions of the kinematic dipole amplitude and direction inferred from samples of $100\,000$ strong lenses with Euclid-quality imaging (blue contours), in combination with spectroscopic stellar velocity dispersion measurements (red contours), or with FP estimates of the velocity dispersion using spectroscopic redshifts (green contours) or photometric redshifts (orange contours). The black lines indicate the true value of $v_{o} = 369.82$\ks, pointing towards Galactic coordinates $(l_o,b_o) = (264.021^\circ,\; 48.253^\circ)$. The contours enclose 39.3\% and 86.5\% of the posterior probability. }
    \label{fig:vel_disp_contour}
\end{figure}

We illustrate the gain in constraining power obtained by adding velocity-dispersion information for the “100k” lens sample relative to the lensing-only case in Figure~\ref{fig:vel_disp_contour}. When only photometric redshift estimates for the lenses and sources are available, and the velocity dispersion 
is inferred using FP relations, the precision on $v_{o}$ improves by a factor of $\sim 3.1$. If spectroscopic redshift measurements for both the lens and the source are available, the constraints improve by a factor of $\sim 3.4$ compared to the lensing-only case. In the perfect scenario in which all lenses have spectroscopic redshifts for both the lens and the source, as well as direct spectroscopic measurements of the stellar velocity dispersion, the constraint improves by a factor of $\sim 8$. In the latter case, the inferred precision becomes sufficient to discriminate between the CMB-inferred value and the value suggested by the source number count analyzes at more than 7$\sigma$. All posterior distributions are reported in Table \ref{tab:posterior_lensing+kinematic} as well as the exclusion significance of the alternative value of $v_{o}$, averaged over ten different realizations of the lens sample. 

Finally, we assess the constraining power of several spectroscopic follow-up scenarios for the Euclid strong-lens sample. In a first, \emph{pessimistic} scenario, we assume that only $15\,000$ lenses receive spectroscopic follow-up, either from large spectroscopic surveys such as DESI \citep{DESI2016} or 4MOST \citep{4MOSToverview, 4SLSLS}, or from dedicated follow-up programs. While both redshift and stellar velocity-dispersion measurements are, in principle, accessible with fiber spectroscopy, the latter are significantly more challenging, as they require a higher signal-to-noise ratio to robustly detect the absorption features of the lens galaxy. As a result, even if a lens is targeted by DESI or 4MOST, a substantial fraction of systems is expected to yield redshift measurements only. In this pessimistic scenario, we therefore assume that velocity-dispersion measurements are successfully obtained for only a third of the spectroscopically targeted lenses, i.e.\,$5\,000$ lenses. The remaining lenses are assumed to have spectroscopic, photometric, or no redshift information, as summarized in Table~\ref{tab:Euclid_scenario}.

In a more \emph{realistic} scenario, we assume that the majority of lenses benefit from at least partial redshift information, either from follow-up spectroscopic observations, Euclid NISP slitless spectroscopy, or, at minimum, reliable photometric redshift estimates. In this case, we adopt a similar success rate for velocity-dispersion measurements, yielding $10\,000$ lenses with kinematic information out of $30\,000$ targeted lenses.

The third, \emph{ideal} scenario corresponds to a dedicated spectroscopic survey of strong lenses, in which all lenses are followed up. We assume a velocity-dispersion measurement success rate of 50\%, resulting in $50\,000$ lenses with stellar velocity-dispersion measurements and an additional $50\,000$ lenses with spectroscopic redshifts for both the lens and the source.

\begin{table}[t]
\centering
\begin{tabular}{lccc}
\hline
& \makecell{Scenario 1 \\ (Pessimistic)} & \makecell{Scenario 2 \\ (Realistic)} & 
\makecell{Scenario 3 \\ (Ideal)}  \\ \hline
No photo-$z$      & 35 000 & - & - \\
FP with photo-$z$   & 50 000 & 70 000 & - \\
FP with spec-$z$    & 10 000 & 20 000 & 50 000\\
With vel.\ disp.  & 5 000 & 10 000 & 50 000\\
\hline
Total            & 100 000 & 100 000 & 100 000\\
\end{tabular}
\caption{Number of lenses with available ancillary information for Euclid pessimistic, realistic and ideal scenarios. \label{tab:Euclid_scenario}}
\end{table}

\begin{table}[t]
\centering
\resizebox{\linewidth}{!}{
\renewcommand{\arraystretch}{1.2}
\begin{tabular}{l|ccc|ccc}

& \multicolumn{3}{c}{$v_{o, \rm true} = 369.82$ \ks} &  \multicolumn{3}{|c}{$v_{o, \rm true} = 1109.46$ \ks} \\ \hline
Lens sample & $v_o$ [km.s$^{-1}$] & $l$ [deg] & $b$ [deg] & $v_o$ [km.s$^{-1}$] & $l$ [deg] & $b$ [deg] \\

\hline
Lensing only 100k & $1057^{+1084}_{-742} \ (0.8\sigma) $ & $233.2^{+53.6}_{-77.3}$ & $4.6^{+29.0}_{-30.7}$ & $1657^{+1183}_{-1058} \ (0.9\sigma) $ & $242.5^{+39.5}_{-47.9}$ & $17.7^{+24.2}_{-21.9}$ \\
FP (Photo-z) 100k & $392^{+331}_{-262} \ (2.6\sigma) $ & $259.5^{+46.9}_{-88.1}$ & $21.8^{+26.9}_{-26.0}$ & $1529^{+992}_{-925} \ (0.9\sigma) $ & $62.0^{+255.3}_{-41.3}$ & $27.3^{+24.4}_{-22.8}$ \\
FP (Spectro-z) 100k & $353^{+307}_{-238} \ (2.7\sigma) $ & $263.2^{+53.3}_{-133.2}$ & $22.9^{+27.9}_{-28.2}$ & $1387^{+1164}_{-940} \ (1.0\sigma) $ & $296.7^{+36.3}_{-147.0}$ & $-11.4^{+24.0}_{-24.9}$ \\
Velocity dispersion 100k & $277^{+111}_{-108} \ (7.5\sigma) $ & $256.8^{+42.0}_{-53.9}$ & $44.0^{+19.5}_{-17.5}$ & $1039^{+109}_{-105} \ (7.4\sigma) $ & $263.1^{+11.2}_{-10.7}$ & $47.7^{+6.3}_{-5.9}$ \\
\hline
Scenario 1 (Pessimistic) & $768^{+220}_{-145} \ (2.5\sigma) $ & $148.5^{+52.0}_{-48.9}$ & $53.9^{+15.9}_{-14.3}$ & $1256^{+384}_{-386} \ (1.8\sigma) $ & $198.7^{+53.3}_{-54.7}$ & $54.1^{+16.4}_{-16.3}$ \\
Scenario 2 (Realistic) & $444^{+283}_{-259} \ (3.4\sigma) $ & $225.2^{+40.4}_{-43.3}$ & $23.1^{+22.6}_{-19.8}$ & $1279^{+361}_{-345} \ (2.3\sigma) $ & $264.4^{+22.7}_{-22.5}$ & $35.7^{+14.2}_{-11.8}$ \\
Scenario 3 (Ideal) & $242^{+146}_{-140} \ (5.1\sigma) $ & $180.1^{+72.1}_{-71.1}$ & $39.6^{+22.6}_{-24.7}$ & $946^{+143}_{-134} \ (5.2\sigma) $ & $243.1^{+21.3}_{-21.4}$ & $55.8^{+10.0}_{-9.0}$ \\

\end{tabular}}

\caption{Posterior constraints on the kinematic dipole amplitude and direction using either lensing-only or combined lensing+kinematic information, obtained from direct spectroscopic measurements or from FP estimates. Scenarios~1 and~2 correspond to plausible ancillary information expected to be available for the Euclid lens sample (see Table~\ref{tab:Euclid_scenario}). The second and third columns correspond to fiducial values of $v_{o}$ equal to the CMB-inferred value and three times the CMB value, respectively. Reported values correspond to the 16$^{\rm th}$, 50$^{\rm th}$, and 84$^{\rm th}$ percentiles. Values in parentheses indicate the one-sided Gaussian-equivalent significance at which the
alternative fiducial value of $v_{o}$ is excluded, averaged over 10 realizations of the lens sample.}
\label{tab:posterior_lensing+kinematic}
\end{table}

\begin{figure}[!]
    \centering
    \includegraphics[width=0.8\linewidth]{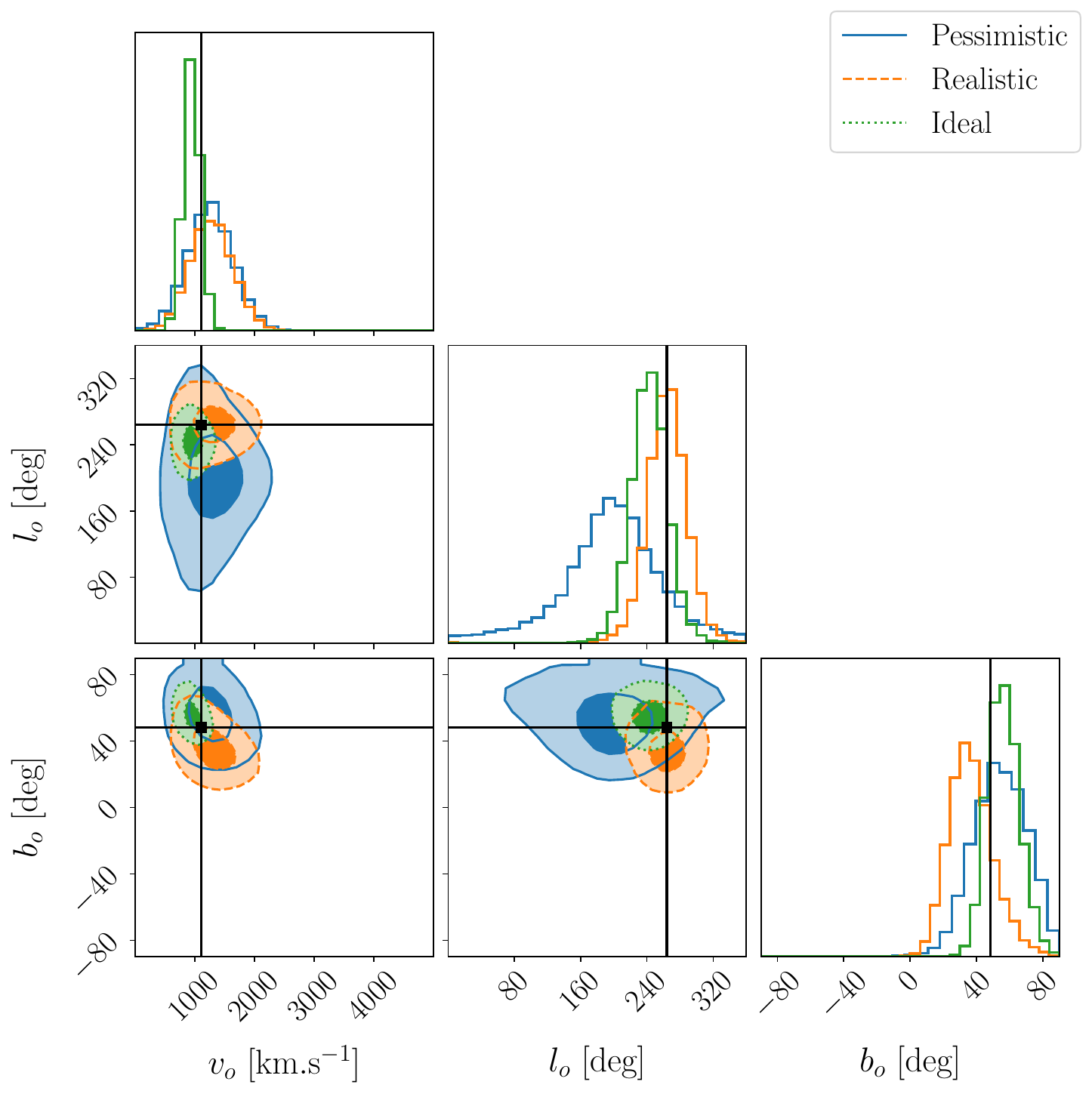}
    \caption{Posterior probability distributions of the kinematic dipole amplitude and direction inferred from samples of $100\,000$ strong lenses with Euclid-quality imaging and different type of ancillary kinematic information, listed in Table \ref{tab:Euclid_scenario}. The black lines indicate the true value of $v_{o} = 1109.46 $\ks, pointing towards Galactic coordinates $(l_o,b_o) = (264.021^\circ,\; 48.253^\circ)$. The contours enclose 39.3\% and 86.5\% of the posterior probability.}
    \label{fig:Euclid_scenario}
\end{figure}

We show the posterior distribution contours for the three scenarios in
Figure~\ref{fig:Euclid_scenario}, assuming a fiducial peculiar velocity
$v_{o,\rm true} = 1109.46$\ks.
The median values, together with the $16^{\rm th}$ and $84^{\rm th}$
percentiles of the marginalized posterior distributions, are reported in Table~\ref{tab:posterior_lensing+kinematic} for a single realization of the lens samples. In addition, we quote the significance at which the alternative value of $v_{o}$ is excluded, averaged over ten independent realizations for each scenario.

For a fiducial value of $v_{o,\rm true} = 1109.46$ \ks, all three spectroscopic follow-up scenarios exclude the CMB-inferred value of $v_{o}$ with a significance ranging from $1.8\sigma$ in the pessimistic scenario to $5.2\sigma$ in the ideal scenario. Conversely, when the CMB-inferred value of $v_{o}$ is adopted as the fiducial velocity, the source number-count value is excluded with a significance ranging from $2.5\sigma$ to $5.1\sigma$ across the same scenarios.
These results indicate that strong-lensing measurements could provide valuable complementary information on the cosmic dipole tension, with a potential discrimination between the two values of $v_{o}$ at the $\sim 2-3\sigma$ level, in a realistic spectroscopic follow-up configuration. In the presence of a dedicated spectroscopic follow-up of strong lenses, however, this probe could make a decisive contribution to this debate, with a discriminating power exceeding $5\sigma$.

\section{Discussion}
\label{sec:discussion}
As shown in the previous section, strong-lensing measurements can reach a sufficient level of precision to discriminate between the CMB-inferred and source number-count values of $v_{o}$, provided that adequate spectroscopic follow-up data are available for a large fraction of the Euclid strong-lens sample. This method would offer an independent test of the cosmic dipole, with systematic uncertainties that are fundamentally different from those affecting source number-count measurements.

\paragraph{Systematic errors:} A primary advantage of this method is that any bias in the absolute calibration of the FP, the measurement of the velocity dispersion, or the dynamical modeling of the galaxies (assumed to be an SIS in this paper) becomes irrelevant, provided such biases are uniform across the sky. 

However, we note several caveats in applying the FP relation to the Euclid lens sample. The FP coefficients we adopt are calibrated on a low-redshift sample ($z\lesssim0.1$; \cite{Said2025}), whereas the lens deflectors span $z\simeq0.1$–$1.5$. The FP zero-point is known to evolve with redshift \citep{Bezanson2013}, to depend on the environment, and the relation does not apply to the $\sim10\%$ of galaxy–galaxy lens deflectors that are late-type galaxies \citep{Bolton2008}. Crucially, however, any such miscalibration that is uniform across the sky, e.g., a mean redshift-dependent offset, the contamination from the minority spiral population, or an environment-dependent normalization, is absorbed by the bias and systematic-error terms ($b_{\rm FP}$, $\sigma_{\rm sys,FP}$), which we fit independently for the photometric- and spectroscopic-redshift subsamples. These effects, therefore, enlarge the systematic-error budget rather than bias the inferred $v_o$. The only component that could bias the dipole is an FP miscalibration that varies coherently across the sky, which will require sophisticated mock simulation that we leave for future work.

In addition, the FP relation is modified by the observer velocity. This effect is directly included in our modeling through two distinct relativistic effects. First, $R_e$ is subject to angular aberration, analogous to that observed in the Einstein angle (Eq.~\eqref{eq:mean_angle_distance_z_main}). Second, the observer’s peculiar velocity affects the conversion of the angular scale $\alpha$ of the galaxy into a physical scale $R_e$ (in kpc) via the angular diameter distance, since the estimated redshift of the galaxy is impacted by the observer’s motion. Both effects are explicitly accounted for in the expression of the likelihood (Eq.~\eqref{eq:logl_probe3}), ensuring that the inferred $v_o$ is unbiased.

Finally, another possible source of systematics is the incomplete sky coverage of the survey. Indeed, since the monopole and the dipole are fitted simultaneously, one could have a bias in the inferred amplitude of the dipole if the sky coverage is insufficient. This is especially true for the lensing-only measurement (Sec.~\ref{sec:Lensing_only}), as the Einstein angle is not compared against any absolute calibration, leading to a potential degeneracy between the monopole and dipole amplitudes if the sky coverage is insufficient. This effect can originate, for example, from galaxy clustering, which can modify both the average Einstein radius of the lenses and their number density on small scales, since denser regions of the Universe are expected to have a higher probability of lensing \citep{Collett2016} and to produce larger Einstein radii. The area covered by the Euclid Wide Survey, which spans approximately $14\,000$ square degrees across both hemispheres, should, however, be largely sufficient to break this degeneracy.

\paragraph{Strong lensing selection function:} One of the main challenges in analyzing large samples of strong lenses is the characterization of the selection function, which can be particularly complex due to the diversity of lens-search algorithms used to identify the systems \citep{Sonnenfeld2022, Sonnenfeld2023, Herle2024}. In Appendix~\ref{app:selection}, we explicitly calculate the impact of the selection function on the dipole, for the lensing only probe. As we will see, the contribution is extremely small, less than $0.1\%$, meaning that it is not necessary to model the selection function to infer the observer's velocity. We consider two selections, the first one on the (magnified) flux of the source, and the second one on the size of the Einstein ring, such that the mean radius is given by
\begin{align}
\label{eq:average_threshold_main}
\langle\theta'_E (\bn') \rangle=&\left(\frac{\dd N(\bn',\theta'_E>\theta'_{E*},F'>F'_{*})}{\dd \Omega'} \right)^{-1}\\
&\times\int \dd z'_s \int \dd z'_l \int_{\theta'_{E*}} \dd\theta'_E \frac{\dd N(\bn',z'_s,z'_l,\theta'_E,F'>F'_{*})}{\dd \Omega' \dd z'_s \dd z'_l \dd \theta'_E} \theta'_E(\bn',z'_s,z'_l)\, ,  \nonumber
\end{align}
where $F'_{*}$ is the observed 
flux threshold of the source and $\theta'_{E*}$ is the observed size threshold. These selection effects have the potential to modulate the mean Einstein radius of the sample $\langle \theta'_E\rangle$ in the dipole direction, leading to a bias in $v_o$ if not properly accounted for. In the presence of kinematic data, however, the Einstein radii are compared on a per-lens basis, and therefore these selection effects, which affect the sample mean, become irrelevant.

Here, we consider hard cuts in flux and Einstein radius, but a smoother selection function would lead to even smaller biases. The flux threshold contributes to the dipole because a fixed observer-frame flux limit, $F'_{*}$, translates into a direction-dependent threshold in the intrinsic luminosity, $L_{*}(\bs{\hat{n}}', z'_s)$. This arises because the luminosity distance—which relates the observed flux to the intrinsic luminosity—is itself affected by the boost. As a consequence, the moving observer does not see the same number of rings in the direction of the boost as in the opposite direction. The same argument applies to the size threshold: a fixed threshold in the observer frame corresponds to different intrinsic sizes of the rings in different directions.

The detailed calculation can be found in Appendix~\ref{app:selection}, where we found\footnote{As shown in the Appendix~\ref{app:selection}, all functions are written in principle in terms of unboosted quantities. However, at the lowest order in $v_o/c$, we can replace the unboosted quantities (a priori unknown) with the boosted ones, such that functions are expressed in terms of observable quantities.} 
\begin{align}
\label{eq:final_threshold}
\langle \theta'_E\rangle&=\langle \theta_E \rangle\left(1-\frac{v_o}{c}\cos\theta'_{\rm cm} \right)\\
&+2\frac{v_o}{c}\cos\theta'_{\rm cm}\int \dd z'_s  \int_{\theta_{E'*}}\dd\theta'_E (\theta'_E-\langle \theta_E\rangle) f_1(z'_s,\theta'_E,F'>F'_*)x(z'_s,\theta'_E,F'_*)\nonumber\\
&+\frac{v_o}{c}\cos\theta'_{\rm cm}(\langle \theta_E\rangle-\theta'_{E*})\int \dd z'_s \int \dd z'_l f_2(z'_s,z'_l,\theta'_{E*},F'>F'_*)\mu(z'_s,z'_l,\theta'_{E*},F'_*)\, .\nonumber
\end{align}
Here $f_1$ is the distribution of detected rings (integrated over the redshift of the lenses) as a function of $\theta'_E$; and $f_2$ is the cumulative distribution above the size threshold, see Eqs.~\eqref{eq:app_f1} and~\eqref{eq:app_f2}. The relativistic magnification bias $x$ and the parameter $\mu$ govern the amplitude of the threshold effects and are given by
\begin{align}
x(z'_s,\theta'_E,F'_*)&=-\frac{\partial\ln}{\partial \ln F'_*}\left(\frac{\dd N(z'_s,\theta'_E,F'>F'_{*})}{\dd \Omega \dd z'_s \dd \theta'_E} \right)\, ,\\
\mu(z'_s,z_l',\theta_{E*},L_*)&= -\frac{\partial\ln}{\partial \ln \theta'_{E*}}\left(\frac{\dd N(z'_s,z'_l,\theta'_E>\theta'_{E*},F'>F'_{*})}{\dd \Omega \dd z'_s \dd z'_l } \right)\, .
\end{align}
They represent the slope of the cumulative distribution of rings above the respective thresholds and therefore quantify how many more systems we detect in one direction because the flux and the size are boosted. For a complete sample, $x$ and $\mu$ vanish exactly. In practice, these parameters can be estimated from the population of rings by binning the systems in flux and size, and measuring the slopes at different thresholds.

What is interesting from Eq.~\eqref{eq:final_threshold} is that the selection effects are strongly suppressed with respect to the case where we simply count objects. This is due to the fact that the mean $\langle\theta'_E\rangle$ is normalized by the number of detected rings. The modulation in the number of rings induced by threshold effects impacts both the numerator and the denominator in Eq.~\eqref{eq:average_threshold_main}, and the two contributions partially cancel out, leading to a negligible impact. For example, the second term in Eq.~\eqref{eq:final_threshold}, representing the effect of the selection in flux, is strongly suppressed by the $(\theta'_E-\langle \theta_E\rangle)$ term. If the parameter $x(z_s',\theta_E',F_*')$ was independent of $z_s'$ and $\theta_E'$, then this contribution would be exactly zero since it would not change the sample mean $\langle \theta_E\rangle$. In practice, this is not the case, since the observed source flux is related to $z_s'$ and $\theta_E'$ through the lensing magnification. However, this means that the surviving contribution is proportional to the change in the mean sample induced by the flux selection. We have computed this effect by evaluating $x$ from our sample and computing the second line of Eq.~\eqref{eq:final_threshold}, and we have found that it generates a negligible contribution to the dipole that is $<10^{-6}$ of the main effect. 

The contribution from the size threshold, encoded in the third line of Eq.~\eqref{eq:final_threshold}, is also suppressed, but this time by a factor $(\langle \theta_E \rangle - \theta'_{E*})$. In this case, there is a balance between this factor and the parameter $\mu$. If most systems have sizes well above the threshold $\theta'_{E*}$, then $(\langle \theta_E \rangle - \theta'_{E*})$ will be large, but the effect will be strongly suppressed by $\mu$, which tends to zero in the limit where there are no rings below the threshold. Conversely, if most rings lie close to the threshold, then $\mu$ can be large, but the effect is suppressed by $(\langle \theta_E \rangle - \theta'_{E*})$, which becomes small for rings near the threshold. We estimate the magnitude of this effect by evaluating $\mu$ from our sample, assuming a threshold of $\theta'_{E*} = 0.2\arcsec$, and computing the third line of Eq.~\eqref{eq:final_threshold}. We find that this contribution is of order $0.1\%$ of the main effect. Although significantly larger than the contribution from the flux threshold, it remains completely negligible compared to the main effect.

This demonstrates that the dipole in the Einstein rings is almost insensitive to the strong lensing selection function, provided it is uniform across the sky. As such, it provides a very robust way of measuring the observer velocity. In other words, neglecting these threshold effects would bias the measurement of the observer velocity by $0.1\%$ at most, largely below the statistical uncertainty.

\paragraph{Limitations of the model:}
In this study, we developed a simple Bayesian framework to assess whether a dipolar modulation of the Einstein angle induced by the observer’s peculiar velocity can be measured from future, yet realistic, samples of strong lenses. Although our lens population model accounts for the selection function of the lenses, we assumed simple Gaussian probability distributions for the measurements of the Einstein angle, stellar velocity dispersion, and redshifts, as well as a trivariate Gaussian model for the FP. However, when propagating these uncertainties through the likelihood (Eq.~\eqref{eq:logl_probe3}), the Gaussian approximation proved insufficient. To address this limitation, we introduced a bias term and an additional systematic error term, which were sufficient to capture the leading-order deviations from Gaussianity and to recover unbiased measurements of $v_{o}$, $l_o$, and $b_o$. A full hierarchical Bayesian inference framework will be required to model these effects more rigorously. Such an approach would allow the measurement uncertainties and intrinsic scatter of the lens and source populations to be modeled consistently at the population level, rather than being absorbed into effective bias and systematic error terms. This may further improve the precision and robustness of the dipole measurement. We leave such an analysis for future work.

Finally, a limitation of this study is that we ignore the impact of ellipticity in the mass distribution of the lensing galaxies. However, we emphasize that we use the \emph{azimuthally averaged} Einstein radius as our primary observable. This quantity is sensitive to the enclosed mass of the galaxy, the angular-diameter distance ratio, and cosmic convergence, which is expected to average out on large scales but is still included as an additional noise contribution to the Einstein radius measurement. In the lensing+kinematic case, ellipticity in the mass distribution of the lens galaxy introduces additional scatter in the predicted central velocity dispersion. Departures from isothermality in the mass profile would produce a similar effect. In this work, we account for both sources of uncertainty by adopting a $5\%$ modeling error in addition to the $5\%$ measurement error. Future analyzes on real data may include full dynamical modeling of the lenses, as well as shear and convergence corrections from weak lensing maps.

\section{Conclusion}
\label{sec:conclusion}
In this work, we present a novel method that could provide a fully-independent way of measuring the kinematic cosmic dipole, based on angle measurement rather than the traditional number counts technique, which relies on flux measurements. Our main findings are the following: 
\begin{itemize}
    \item A detection of the dipolar modulation of the Einstein radii of strong lenses is possible using the lensing information contained in Euclid images alone, but only in the case where the velocity is as large as inferred from the number-count analyses.
    \item The inclusion of kinematic information, either through spectroscopic velocity-dispersion measurements or via FP relations, significantly improves the constraining power of the method. Even in the absence of spectroscopic redshift information, the combination of FP relations and photometric redshifts leads to an improvement by a factor of $\sim 3$ compared to lensing-only constraints.
    \item Already existing or planned spectroscopic follow-up observations from 4MOST \cite{4SLSLS} and DESI \cite[e.g.,][]{Huang2025} should yield a discriminating power at the level of $2.3\sigma$ to $ 3.4\sigma$ between the value of $v_{o}$ measured from the CMB and that measured from source number counts.
    \item A hypothetical dedicated strong-lensing spectroscopic follow-up of the full Euclid sample could provide a decisive contribution to this debate, with a discriminating power reaching $\sim 5\sigma$.
    \item The method is mostly insensitive to the lens selection function, provided it is spatially uniform over the survey footprint. Similarly, any bias in the lens mass modeling or in the dynamical modeling of the lenses can be self-calibrated from the data and does not affect the measurement of $v_{o}$, as long as these biases are constant across the sky.
\end{itemize}
The recent discovery of hundreds of new strong-lens systems in the Euclid Q1 data highlights the potential to increase the current sample of known lenses by two orders of magnitude with the full Euclid survey. With such a large sample, we propose a direct cosmological test that can be performed using the imaging survey data, complemented by modest spectroscopic follow-up from Euclid NISP, DESI, or 4MOST.

\section*{Acknowledgments}
We are grateful to Koen Kuijken for identifying a calculation error in an earlier version of this manuscript. MM acknowledges support from the Swiss National Science Foundation (SNSF) through return CH grant P5R5PT\_225598 and Ambizione grant PZ00P2\_223738. This work has received funding from the European Research Council (ERC) under the European Union's Horizon 2020 research and innovation program (LensEra: grant agreement No 945536). TEC is funded by the Royal Society through a University Research Fellowship. CB acknowledges support from the Swiss National Science Foundation (SNSF).

\appendix

\section{Calculation of the mean Einstein radius integrated over redshifts}
\label{app:redshift}

To calculate Eq.~\eqref{eq:average_redshift}, we need to compute the Einstein angle and the number of sources as functions of the measured redshifts $z'_s$ and $z'_l$. These redshifts are affected by the observer velocity $v_o$ through
\begin{align}
1+z'_s&=(1+z_s)\left(1-\frac{v_o}{c}\cos\theta'_{\rm cm}\right)\, ,\\
1+z'_l&=(1+z_l)\left(1-\frac{v_o}{c}\cos\theta'_{\rm cm}\right)\, .
\end{align}
The Einstein radius $\theta'_E$ at fixed $z'_s$ and $z'_l$ is given by
\begin{align}
\label{eq:theta_z_1}
\theta'_{E}(\delta',\bn',z'_s,z'_l)= \theta'_{E}(\delta',\bn',r'(z'_s,\bn'),r'(z'_l,\bn'))\, ,  
\end{align}
where $r'(z'_s,\bn')$ is the comoving distance associated with the boosted redshift $z'_s$, which depends on direction $\bn'$ through the impact of the boost on the redshift. We split this distance into an isotropic part and a perturbation
\begin{align}
r'(z'_s,\bn')=r(z'_s)+\delta r(z'_s,\bn')\, .     
\end{align}
As shown in~\cite{Dalang:2021ruy}, the perturbation in the distance is given by 
\begin{align}
\delta r(z'_s,\bn')=\frac{1}{\HH(z'_s)}\frac{v_o}{c}\cos\theta'_{\rm cm}\, .  
\end{align}
Doing the same split for the redshift of the lens, we can insert it into~\eqref{eq:theta_z_1} and Taylor expand at linear order in $v_o/c$
\begin{align}
&\theta'_{E}(\bn',z'_s,z'_l)= \theta'_{E}(r(z'_s),r(z'_l))+\left(\frac{\partial\theta_E}{\partial r_s\HH(z'_s)}+\frac{\partial\theta_E}{\partial r_l\HH(z'_l)} \right)\frac{v_o}{c}\cos\theta'_{\rm cm}\nonumber\\
&=\theta_{E}(r(z'_s),r(z'_l))\left(1-\frac{v_o}{c}\cos\theta'_{\rm cm}\right)
+\left((1+z_z)\frac{\partial\theta_E}{\partial z_s}+(1+z_l)\frac{\partial\theta_E}{\partial z_l} \right)\frac{v_o}{c}\cos\theta'_{\rm cm}\, , \label{eq:theta_z}
\end{align}
where in the second line we have used that $\partial/\partial r=\HH(1+z)\partial/\partial z$.
From this, we see that the fact that the redshifts of the lens and the source are affected by the boost adds another contribution to $\theta'_{E}$. Physically, this is  due to the fact that a source at a fixed redshift $z'_s$ is not at the same physical distance from the observer if it is in the direction of the boost or if it is in the opposite direction. Since the Einstein radius depends on the distance of the source, this impacts the observed $\theta'_{E}$. 

To obtain the mean Einstein radius in a solid angle $\dd\Omega'$, we need to weight it by the number of systems at fixed $z'_s, z'_l$. We have
\begin{align}
\frac{\dd N(z'_s,z'_l,\bn')}{\dd \Omega'\dd z'_s \dd z'_l}&= \frac{\dd N(r(z'_s),r(z'_l))}{\dd \Omega \dd z_s \dd z_l}\left(1-\frac{\delta\Omega}{\Omega}-\frac{\delta z_s}{1+z_s}-\frac{\delta z_l}{1+z_l} \right) \nonumber \\
&+\frac{\partial}{\partial r_s}\left(\frac{\dd N}{\dd \Omega \dd z_s \dd z_l} \right) \delta r_s+\frac{\partial}{\partial r_l}\left(\frac{\dd N}{\dd \Omega \dd z_s \dd z_l} \right) \delta r_l\nonumber\\
&= \frac{\dd N(r(z'_s),r(z'_l))}{\dd \Omega \dd z_s \dd z_l}\left(1 + 4 \frac{v_o}{c}\cos\theta'_{\rm cm} \right) \nonumber \\
&+\left[(1+z_s)\frac{\partial}{\partial z_s}\left(\frac{\dd N}{\dd \Omega \dd z_s \dd z_l} \right) +(1+z_l)\frac{\partial}{\partial z_l}\left(\frac{\dd N}{\dd \Omega \dd z_s \dd z_l} \right)\right] \frac{v_o}{c}\cos\theta'_{\rm cm}\, .\label{eq:N_z}
\end{align}
Multiplying Eq.~\eqref{eq:N_z} with~\eqref{eq:theta_z} and using that
\begin{align}
\frac{\partial}{\partial z_s}\left(\frac{\dd N}{\dd \Omega \dd z_s \dd z_l} \right) (1+z_s)+ \left[\frac{\partial\theta_E}{\partial r_s\HH(z'_s)}+1\right]\frac{\dd N}{\dd \Omega \dd z_s \dd z_l}= \frac{\partial}{\partial z_s}\left[\frac{\dd N}{\dd \Omega \dd z_s \dd z_l}(1+z_s)\theta_E \right]
\end{align}
we obtain
\begin{align}
&\int \dd z'_s\int \dd z'_l \theta'_E(\bn',z_s',z'_l) \frac{\dd N(z'_s,z'_l,\bn')}{\dd \Omega'\dd z'_s \dd z'_l}\label{eq:app_num}\\
&=\int \dd z'_s\int \dd z'_l\frac{\dd N(r(z'_s),r(z'_l))}{\dd \Omega \dd z_s \dd z_l}\theta_E\left(1+2\frac{v_o}{c}\cos\theta'_{\rm cm}-\frac{v_o}{c}\cos\theta'_{\rm cm} \right)\nonumber\\
&+\int \dd z'_s \int \dd z'_l\left\{\frac{\partial}{\partial z_s}\left[\frac{\dd N}{\dd \Omega \dd z_s \dd z_l}(1+z_s)\theta_E \right]+\frac{\partial}{\partial z_l}\left[\frac{\dd N}{\dd \Omega \dd z_s \dd z_l}(1+z_l)\theta_E \right]\right\}\frac{v_o}{c}\cos\theta'_{\rm cm}\, .\nonumber
\end{align}
The terms in the last line can be directly integrated, giving rise to boundary terms that exactly vanish since the number of rings at redshift 0 and at infinity is zero. The only effect that survives is therefore the effect of aberration, which modifies both the number of rings (second term in the second line of \eqref{eq:app_num}) and the Einstein radius (third term in the second line of \eqref{eq:app_num}). A similar calculation can be performed for the denominator of  Eq.~\eqref{eq:average_redshift}, $dN(\bn')/d\Omega'$, leading to
\begin{align}
\frac{\dd N(\bn')}{\dd \Omega'}= \int \dd z'_s\int \dd z'_l \frac{\dd N(r(z'_s),r(z'_l))}{\dd \Omega \dd z_s \dd z_l}\theta_E\left(1+2\frac{v_o}{c}\cos\theta'_{\rm cm}\right)\, .\label{eq:app_N}
\end{align}
Taking the ratio of Eqs.~\eqref{eq:app_num} and~\eqref{eq:app_N} , we obtain the mean
\begin{align}
\label{eq:mean_angle_distance_z}
\langle\theta'_E (\bn') \rangle=\left(1 - \frac{v_o}{c}\cos\theta'_{\rm cm}\right)\int \dd z'_s \int \dd z'_l f(z'_s,z'_l) \theta_E(z'_s,z'_l)\, , \end{align}
with $f(z'_s,z'_l)$ given in Eq.~\eqref{eq:f_z}.

\section{Impact of the selection function on the dipole}
\label{app:selection}

We compute here the impact of selection in flux and size on the dipole. We work in redshift space, but as we will see, we would obtain the same result using comoving distances once the signal is integrated from 0 to $\infty$. 

The mean Einstein radius, averaged over sources with a flux larger than $F'_{*}$ and a size larger than $\theta'_{E*}$, is given by
\begin{align}
\label{eq:average_threshold}
\langle\theta'_E (\bn') \rangle=&\left(\frac{\dd N(\bn',\theta'_E>\theta'_{E*},F'>F'_{*})}{\dd \Omega'} \right)^{-1}\\
&\times\int \dd z'_s \int \dd z'_l \int_{\theta'_{E*}} \dd\theta'_E \frac{\dd N(\bn',z'_s,z'_l,\theta'_E,F'>F'_{*})}{\dd \Omega' \dd z'_s \dd z'_l \dd \theta'_E} \theta'_E(\bn',z'_s,z'_l)\, .  \nonumber
\end{align}
In principle, we could also add a threshold on the flux of the lens, but this is irrelevant since the vast majority of sources for which the ring is detected have a lens that is also detected. 

The observed flux is related to the intrinsic source luminosity $L$ through the luminosity distance
\begin{align}
F'=\frac{L}{4\pi d^{'2}_L} \, .  
\end{align}
The observed flux threshold $F_*'$ is, by construction, independent of direction. However, since the luminosity distance depends on direction through the boost, the intrinsic luminosity threshold also depends on direction
\begin{align}
\label{eq:luminosity}
L_*(r(z_s'))+\delta L_*(\bn',z'_s)=4\pi d_L^2\left(1+2\frac{\delta d_L}{d_L} \right)F_*'=4\pi d_L^2\left(1+2\frac{v_o\cos\theta'_{\rm cm}}{c\HH r} \right)F_*'\, , 
\end{align}
where the luminosity distance fluctuation has been computed in~\cite{Bonvin:2005ps}. We will therefore detect a different number of rings above the flux threshold in the direction of the boost than in the opposite direction. If this effect were independent of the distance, it would affect the numerator and the denominator of~\eqref{eq:average_threshold} in the same way, and it would cancel. However, as we will see, this is not the case, and there is a non-zero contribution.

Similarly to the flux, we can relate the observed Einstein radius to its intrinsic size (which depends only on the mass of the lens) by factoring out the dependence on angular diameter distances
\begin{align}
\theta_E'=\theta^{\rm  int}_E \left(1-\frac{v_o}{c}\cos\theta'_{\rm cm}\right)g(z_s',z'_l) \, , 
\end{align}
where $g(z'_s,z'_l)= d_{ls}(z'_s,z'_l)/d_s(z'_s)$. 
As for the luminosity, since the observed size threshold is independent of direction, the intrinsic size threshold acquires a dependence on $\bn'$ through the boost
\begin{align}
\label{eq:intsize}
\theta^{\rm int}_{E*}(r(z'_s),r(z'_l))+\delta\theta^{\rm int}_{E*}(\bn',z'_s, z'_l)=\frac{\theta'_{E*}}{g(r(z_s'),r(z'_l))} \left(1+\frac{v_o}{c}\cos\theta'_{\rm cm}-\frac{\partial g}{\partial  r_s}\frac{\delta r_s}{g}-\frac{\partial g}{\partial  r_l}\frac{\delta r_l}{g}\right) \, .
\end{align}
We then have 
\begin{align}
&\int_{\theta'_{E*}}\dd \theta'_E\frac{\dd N(\bn',z'_s,z'_l,\theta'_E,F'>F'_{*})}{\dd \Omega' \dd z'_s \dd z'_l \dd \theta'_E}\theta_E'=
\int_{\thint_{E*}+\delta\thint_{E*}}\dd \theta^{\rm int}_E\frac{\dd N(\bn',z'_s,z'_l,\theta^{\rm int}_E,L>L_{*}+\delta L_*)}{\dd \Omega' \dd z'_s \dd z'_l \dd \theta^{\rm  int}_E}\theta_E'\nonumber\\
&=\int_{\thint_{E*}}\dd\thint_E \thint_E\Bigg\{\frac{\dd N(r(z'_s),r(z'_l),\thint_E,L>L_{*})}{\dd \Omega \dd z_s \dd z_l \dd \thint_E}g(r(z'_s),r(z'_l))\left(1+\frac{v_o}{c}\cos\theta'_{\rm cm} \right)\nonumber\\
&\quad+\frac{\partial}{\partial z'_s}\left[\frac{\dd N(r(z'_s),r(z'_l),\thint_E,L>L_{*})}{\dd \Omega \dd z_s \dd z_l \dd \thint_E}(1+z'_s)g(r(z'_s),r(z'_l)) \right]\frac{v_o}{c}\cos\theta'_{\rm cm}\nonumber\\
&\quad+\frac{\partial}{\partial z'_l}\left[\frac{\dd N(r(z'_s),r(z'_l),\thint_E,L>L_{*})}{\dd \Omega \dd z_s \dd z_l \dd \thint_E}(1+z'_l)g(r(z'_s),r(z'_l)) \right]\frac{v_o}{c}\cos\theta'_{\rm cm}\nonumber\\
&\quad+\frac{\partial}{\partial L_*}\left[\frac{\dd N(r(z'_s),r(z'_l),\thint_E,L>L_{*})}{\dd \Omega \dd z_s \dd z_l \dd \thint_E}\right]g(r(z'_s),r(z'_l))\delta L_*\Bigg\}\nonumber\\
&+\frac{\partial}{\partial \thint_{E*}}\left[\int_{\thint_{E*}}\dd\thint_{E}\thint_{E}\frac{\dd N(r(z'_s),r(z'_l),\thint_E,L>L_{*})}{\dd \Omega \dd z_s \dd z_l \dd \thint_E}g(r(z'_s),r(z'_l)) \right]\delta\thint_{E*}\, .
\label{eq:inttheta}
\end{align}

As in the case without threshold effects in Appendix~\ref{app:redshift}, we want to group the terms in a total derivative that vanishes at the boundary. The difference here is that both $L_*$ and $\thint_{E*}$ depend on redshifts. $L_*$ depends only on the redshift of the source, while $\thint_{E*}$ depends on both redshifts. As a consequence, the total derivative acquires additional terms. More precisely the total derivative over $z_s'$ is given by
\begin{align}
\label{eq:total_dz}
&\frac{d}{dz'_s}\Bigg\{\int_{\thint_{E*}}\dd\thint_E \thint_E \frac{\dd N(r(z'_s),r(z'_l),\thint_E,L>L_{*})}{\dd \Omega \dd z_s \dd z_l \dd \thint_E}(1+z'_s)g(r(z'_s),r(z'_l))\Bigg\} \frac{v_o}{c}\cos\theta'_{\rm cm}\\
&=\int_{\thint_{E*}}\dd\thint_E \thint_E\Bigg\{\frac{\partial}{\partial z'_s}\left[\frac{\dd N(r(z'_s),r(z'_l),\thint_E,L>L_{*})}{\dd \Omega \dd z_s \dd z_l \dd \thint_E}(1+z'_s)g(r(z'_s),r(z'_l)) \right]\nonumber\\
&+\frac{\partial}{\partial L_*}\left[\frac{\dd N(r(z'_s),r(z'_l),\thint_E,L>L_{*})}{\dd \Omega \dd z_s \dd z_l \dd \thint_E}\right]g(r(z'_s),r(z'_l))(1+z'_s)\frac{\dd L_*}{\dd z'_s}\Bigg\}\frac{v_o}{c}\cos\theta'_{\rm cm}\nonumber\\
&+\frac{\partial}{\partial \thint_{E*}}\left[\int_{\thint_{E*}}\dd\thint_{E}\thint_{E}\frac{\dd N(r(z'_s),r(z'_l),\thint_E,L>L_{*})}{\dd \Omega \dd z_s \dd z_l \dd \thint_E}\right]g(r(z'_s),r(z'_l))(1+z'_s)\frac{\dd\thint_{E*}}{\dd z'_s} \frac{v_o}{c}\cos\theta'_{\rm cm}\nonumber\, .
\end{align}
A similar expression can be derived for $z'_l$, without the contribution from the third line since $L_*$, which is the intrinsic luminosity of the source, does not depend on the redshift of the lens. Using Eq.~\eqref{eq:total_dz} in Eq.~\eqref{eq:inttheta} and integrating over redshifts, we find
\begin{align}
&\int \dd z'_s \int \dd z'_l \int_{\theta'_{E*}} \dd\theta'_E \frac{\dd N(\bn',z'_s,z'_l,\theta'_E,F'>F'_{*})}{\dd \Omega' \dd z'_s \dd z'_l \dd \theta'_E} \theta'_E(\bn',z'_s,z'_l) \\
&=\int \dd z'_s \int \dd z'_l\Bigg\{\int_{\thint_{E*}}\dd\thint_E \thint_E\Bigg[\frac{\dd N(r(z'_s),r(z'_l),\thint_E,L>L_{*})}{\dd \Omega \dd z_s \dd z_l \dd \thint_E}g(r(z'_s),r(z'_l))\left(1+\frac{v_o}{c}\cos\theta'_{\rm cm} \right)\nonumber\\
&+\frac{\partial}{\partial L_*}\left(\frac{\dd N(r(z'_s),r(z'_l),\thint_E,L>L_{*})}{\dd \Omega \dd z_s \dd z_l \dd \thint_E}\right)g(r(z'_s),r(z'_l))\left(\delta L_*- (1+z'_s)\frac{\dd L_*}{\dd z'_s}\frac{v_o}{c}\cos\theta'_{\rm cm}\right)\Bigg]\nonumber\\
&+\frac{\partial}{\partial \thint_{E*}}\left(\int_{\thint_{E*}}\dd\thint_{E}\thint_{E}\frac{\dd N(r(z'_s),r(z'_l),\thint_E,L>L_{*})}{\dd \Omega \dd z_s \dd z_l \dd \thint_E}\right)\nonumber\\
&\quad \times g(r(z'_s),r(z'_l))\left[\delta\thint_{E*}- \left((1+z'_s)\frac{\dd\thint_{E*}}{\dd z'_s} +(1+z'_l)\frac{\dd\thint_{E*}}{\dd z'_l}\right)\frac{v_o}{c}\cos\theta'_{\rm cm} \right]\Bigg\}\, ,\nonumber
\end{align}
where we have used that the total derivatives over $z'_s$ and $z'_l$ cancel when integrated over all redshifts, since the number of rings at redshift 0 and infinity is zero.
The luminosity perturbation and its redshift derivative can be directly computed from Eq.~\eqref{eq:luminosity}
\begin{align}
\delta L_*&=\frac{2L_*}{\HH r}  \frac{v_o}{c}\cos\theta'_{\rm cm}\, , \\
(1+z'_s)\frac{\dd L_*}{\dd z'_s}&=2L_*\left(1+\frac{1}{r_s\HH}\right)\, .
\end{align}
Similarly, the intrinsic size perturbation and its redshift derivatives can be computed from Eq.~\eqref{eq:intsize}
\begin{align}
&\delta\thint_{E*}=\thint_{E*}\left(\frac{v_o}{c}\cos\theta'_{\rm cm}-\frac{\partial g}{\partial r_s}\frac{\delta r_s}{g}-\frac{\partial g}{\partial r_l}\frac{\delta r_l}{g}\right)  \, , \\
&(1+z'_s)\frac{\dd \thint_{E*}}{\dd z'_s}+(1+z'_l)\frac{\dd \thint_{E*}}{\dd z'_l}=-\frac{\thint_{E*}}{g}\left(\frac{\partial g}{\partial z'_s}+\frac{\partial g}{\partial z'_l}\right)\, .
\end{align}
With this, Eq.~\eqref{eq:total_dz} becomes
\begin{align}
\label{eq:final_size}
&\int \dd z'_s \int \dd z'_l \int_{\theta'_{E*}} \dd\theta'_E \frac{\dd N(\bn',z'_s,z'_l,\theta'_E,F'>F'_{*})}{\dd \Omega' \dd z'_s \dd z'_l \dd \theta'_E} \theta'_E(\bn',z'_s,z'_l) \\
&=\int \dd z'_s \int \dd z'_l\int_{\thint_{E*}}\dd\thint_E \thint_E\Bigg[\frac{\dd N(r(z'_s),r(z'_l),\thint_E,L>L_{*})}{\dd \Omega \dd z_s \dd z_l \dd \thint_E}g(r(z'_s),r(z'_l))\left(1+\frac{v_o}{c}\cos\theta'_{\rm cm} \right)\nonumber\\
&-2\frac{\partial}{\partial L_*}\left(\frac{\dd N(r(z'_s),r(z'_l),\thint_E,L>L_{*})}{\dd \Omega \dd z_s \dd z_l \dd \thint_E}\right)g(r(z'_s),r(z'_l))L_*\frac{v_o}{c}\cos\theta'_{\rm cm}\Bigg]\nonumber\\
&-\int \dd z'_s \int \dd z'_l\,\thint_{E*}\frac{\dd N(r(z'_s),r(z'_l),\thint_{E*},L>L_{*})}{\dd \Omega \dd z_s \dd z_l \dd \thint_E}
g(r(z'_s),r(z'_l))\thint_{E*}\frac{v_o}{2c}\cos\theta'_{\rm cm}\, .\nonumber
\end{align}
The last two lines are new contributions generated by the threshold effects: since the boost impacts the observed flux and the observed size of the Einstein ring, we see a different number of events in the direction of the boost and in the opposite direction, which impacts the integral over redshifts.

To obtain the mean Einstein radius, we need, however, to divide Eq.~\eqref{eq:final_size} by the number of events observed in the direction $\bn'$. Following the same steps as for Eq.~\eqref{eq:final_size}, we find
\begin{align}
\label{eq:final_N}
&\frac{\dd N(\bn',\theta'_E>\theta'_{E*},F'>F'_{*})}{\dd \Omega'}\\
&= \int \dd z'_s \int \dd z'_l\int_{\thint_{E*}}\dd\thint_E \Bigg[\frac{\dd N(r(z'_s),r(z'_l),\thint_E,L>L_{*})}{\dd \Omega \dd z_s \dd z_l \dd \thint_E}\left(1+2\frac{v_o}{c}\cos\theta'_{\rm cm} \right)\nonumber\\
&-2\frac{\partial}{\partial L_*}\left(\frac{\dd N(r(z'_s),r(z'_l),\thint_E,L>L_{*})}{\dd \Omega \dd z_s \dd z_l \dd \thint_E}\right)L_*\frac{v_o}{c}\cos\theta'_{\rm cm}\Bigg]\nonumber\\
&-\int \dd z'_s \int \dd z'_l\frac{\dd N(r(z'_s),r(z'_l),\thint_{E*},L>L_{*})}{\dd \Omega \dd z_s \dd z_l \dd \thint_E}
\thint_{E*}\frac{v_o}{2c}\cos\theta'_{\rm cm}\, . \nonumber 
\end{align}
We then rewrite Eqs.~\eqref{eq:final_size} and~\eqref{eq:final_N} in terms of the Einstein radius $\theta_E=g(r(z_s',r(z_l')))\thint_E $, and we take the ratio. This gives the final result
\begin{align}
\label{eq:final_threshold_app}
\langle \theta'_E\rangle&=\langle \theta_E \rangle\left(1-\frac{v_o}{c}\cos\theta'_{\rm cm} \right)\\
&+2\frac{v_o}{c}\cos\theta'_{\rm cm}\int \dd z'_s  \int_{\theta_{E*}}\dd\theta_E (\theta_E-\langle \theta_E\rangle) f_1(z'_s,\theta_E,L>L_*)x(z'_s,\theta_E,L_*)\nonumber\\
&+\frac{v_o}{c}\cos\theta'_{\rm cm}(\langle \theta_E\rangle-\theta_{E*})\int \dd z'_s \int \dd z'_l f_2(z'_s,z'_l,\theta_E>\theta_{E*},L>L_*)\mu(z'_s,z'_l,\theta_{E*},L_*)\, .\nonumber
\end{align}

Here $f_1$ is the distribution of systems, already integrated over the redshifts of the lenses (which do not impact the flux threshold)  
\begin{align}
\label{eq:app_f1}
f_1(z'_s,\theta_E,L>L_*) =\left(\frac{dN(\theta_E>\theta_{E*},L>L_*)}{d\Omega} \right)^{-1}\frac{\dd N(r(z'_s),\theta_E,L>L_{*})}{\dd \Omega \dd z_s \dd \theta_E}\, ,
\end{align}
and $f_2$ is the cumulative distribution with sizes above the threshold
\begin{align}
\label{eq:app_f2}
f_2(z'_s, z'_l,\theta_E>\theta_{E*},L>L_*) =\left(\frac{dN(\theta_E>\theta_{E*},L>L_*)}{d\Omega} \right)^{-1}\frac{\dd N(r(z'_s),r(z'_l),\theta_E>\theta_{E*},L>L_{*})}{\dd \Omega \dd z_s \dd}\, .
\end{align}
These two distributions are unboosted quantities, which depend on the unboosted distance $r$, the unboosted angle $\theta_E$, and the intrinsic unboosted luminosity threshold $L_*$. In practice, however, to predict the amplitude of the second and third lines in Eq.~\eqref{eq:final_threshold_app}, we can directly use the measured mean redshift distribution as a function of the observed redshifts, the observed angle $\theta'_E$, and the observed flux threshold $F'>F'_*$, since $f_1$ and $f_2$ already multiply $v_o/c$.

The second line in Eq.~\eqref{eq:final_threshold_app} encodes the impact of the flux threshold on the mean Einstein ring. It depends on the so-called (relativistic) magnification bias, i.e., the slope of the cumulative number of systems with a flux larger than the flux threshold
\begin{align}
x(z'_s,\theta_E,L_*)\equiv -\frac{\partial\ln}{\partial \ln L_*}\left(\frac{\dd N(r(z'_s),\theta_E,L>L_{*})}{\dd \Omega \dd z_s  \dd \theta_E} \right)   =-\frac{\partial\ln}{\partial \ln F'_*}\left(\frac{\dd N(z'_s,\theta'_E,F'>F'_{*})}{\dd \Omega \dd z'_s \dd \theta'_E} \right)\, ,
\end{align}
where in the second equality we have used the fact that $x$ already multiplies $v_o/c$. 

Finally, the third line in Eq.~\eqref{eq:final_threshold_app} encodes the impact of the size threshold. It depends on the slope of the cumulative number of systems with a size larger than $\theta_{E*}$
\begin{align}
\mu(z'_s,z_l',\theta_{E*},L_*)\equiv -\frac{\partial\ln}{\partial \ln \theta_{E*}}\left(\frac{\dd N(r(z'_s),r(z'_l),\theta_E>\theta_{E*},L>L_{*})}{\dd \Omega \dd z_s \dd z_l } \right)\, .
\end{align}
As for the magnification bias, unboosted quantities can be replaced by the measured ones in the definition of $\mu$.

\section{Simulation of the Fundamental Plane observables}
\label{ap:A}
We define the FP logarithmic variables: 
\begin{equation}
\bs{x} \equiv (r,\, s,\, i) \equiv 
\left(\log R_e,\ \log \sigma_{\mathrm{FP}},\ \log I_e \right),
\end{equation}
and we model the intrinsic distribution of galaxies as a tri-variate Gaussian, following e.g. \cite{Colless2001b}: 

\begin{equation}
P(\bs{x}) =
\mathcal{N}(\boldsymbol{\mu},\,\boldsymbol{\Sigma})\,,
\end{equation}
with mean
\begin{equation}
\label{eq:FP_mean}
\boldsymbol{\mu} =
(\bar r,\ \bar s,\ \bar i)\,,
\end{equation}
and covariance matrix $\boldsymbol{\Sigma}$.

The covariance matrix is constructed from the orthonormal eigenbasis:
\begin{equation}
\{\hat{\bs{u}}_1,\ \hat{\bs{u}}_2,\ \hat{\bs{n}}\},
\end{equation}
where $\hat{\bs{u}}_1$ and $\hat{\bs{u}}_2$ span the FP and
$\hat{\bs{n}}$ is perpendicular to it. The unit vector perpendicular to the FP, $\hat{\bs{n}}$, can be related to the cartesian basis ($\bs{\hat{e}}_r, \bs{\hat{e}}_s, \bs{\hat{e}}_i$) : 

\begin{equation}
\hat{\bs{n}} =
\frac{\hat{\bs e}_r-a\hat{\bs e}_s-b\hat{\bs e}_i}{\sqrt{1+a^2+b^2}}. 
\end{equation}
In this basis, the intrinsic dispersions are $(\sigma_1,\ \sigma_2,\ \sigma_3)$, with $\sigma_1$ being the thickness of the FP. The covariance matrix is then
\begin{equation}
\label{eq:Sigma_rsi}
\boldsymbol{\Sigma}_{rsi}
=
\bs{V}
\begin{pmatrix}
\sigma_2^2 & 0 & 0 \\
0 & \sigma_3^2 & 0 \\
0 & 0 & \sigma_1^2
\end{pmatrix}
\bs{V}^{\mathsf T}\,,
\end{equation}
where $\bs{V} = (\hat{\bs{u}}_1,\hat{\bs{u}}_2,\hat{\bs{n}})$. For each lens of our sample, the conditional distribution of $(r,i)$, given the true value $s$, remains Gaussian:
\begin{equation}
P(r,i \mid s) =
\mathcal{N}(\boldsymbol{\mu}_{RI|s},\ \boldsymbol{\Sigma}_{RI|s})\,,
\end{equation}
with conditional mean
\begin{equation}
\boldsymbol{\mu}_{ri|s}
=
\boldsymbol{\mu}_{ri}
+
\boldsymbol{\Sigma}_{ri,s}\,
\Sigma_{ss}^{-1}
\,(s - \bar s)\,,
\end{equation}
and conditional covariance
\begin{equation}
\boldsymbol{\Sigma}_{RI|s}
=
\boldsymbol{\Sigma}_{RI}
-
\boldsymbol{\Sigma}_{RI,s}\,
\Sigma_{ss}^{-1}
\boldsymbol{\Sigma}_{s,RI}, 
\end{equation}
where $\boldsymbol{\Sigma}_{RI}$ is the $2\times2$ covariance of $(r,i)$, $\boldsymbol{\Sigma}_{RI,s}$ is the covariance between $(r,i)$ and $s$, and $\Sigma_{ss}$ is the variance of $s$

We emulate the measurement errors $(\delta r, \delta i)$ on the observable $(r,i)$ for each lens system from a multivariate Gaussian distribution, following \cite{Said2020}: 

\begin{align}
\label{eq:FP_error_matrix}
\begin{pmatrix}\delta r \\[2pt] \delta i\end{pmatrix}
&\sim \mathcal{N}\!\left(0, \bs{E}_n \;
\right), \qquad 
\bs{E}_n =
\begin{pmatrix}
\epsilon_r^2 & \rho_{ri}\,\epsilon_r\,\epsilon_i \\
\rho_{ri}\,\epsilon_r\,\epsilon_i & \epsilon_i^2
\end{pmatrix},
\end{align}
where $\rho_{ri} = -0.95$ is the correlation between the
photometric quantities $r$ and $i$. $\epsilon_r$ and $\epsilon_i$ were found to be 0.049 dex (11\%) and 0.073 dex (17\%) in \cite{Magoulas2012}, from ground-based imaging. Although these could be substantially improved with Euclid \citep{EuclidBretonniere2023}, we adopt a conservative value of $\epsilon_r=0.041$ dex (10\%) and $\epsilon_i=0.06$ dex (15\%).

The observed quantities are
\begin{align}
r_{\rm obs} &= r_{\rm true} + \delta r, &
i_{\rm obs} &= i_{\rm true} + \delta i.
\end{align}

Adopting the same tri-variate Gaussian model and defining the total covariance matrix $\bs{C}_n = \boldsymbol{\Sigma} + \bs{E}_n$. We can obtain an estimate of $s$ given the observed ($r_{\rm obs}, i_{\rm obs}$)

\begin{equation}
P(s \mid r,i)
=
\mathcal{N}\!\left(\mu_{s|RI},\,\sigma^2_{s|RI}\right),
\end{equation}
with conditional mean
\begin{equation}
\mu_{s|RI}
=
\mu_s
+
\bs{C}_{s,RI}\,
\bs{C}_{RI}^{-1}
\left(
\begin{pmatrix}
r \\ i
\end{pmatrix}
-
\boldsymbol{\mu}_{RI}
\right)
\end{equation}
and conditional variance
\begin{equation}
\sigma^2_{s|RI}
=
C_{ss}
-
\bs{C}_{s,RI}\,
\bs{C}_{RI}^{-1}
\bs{C}_{RI,s},
\end{equation}

where we partitioned the mean vector and covariance matrix as follows
\begin{equation}
\boldsymbol{\mu} =
\begin{pmatrix}
\boldsymbol{\mu}_{RI} \\
\mu_s
\end{pmatrix},
\qquad
\bs{C}_n =
\begin{pmatrix}
\bs{C}_{RI} & \bs{C}_{RI,s} \\
\bs{C}_{s,RI} & C_{ss}
\end{pmatrix},
\qquad
\boldsymbol{\mu}_{RI} = (\bar r,\bar i). \\
\end{equation}

\section{Observer velocity effect on the Fundamental Plane observables}
\label{ap:B}

Here we describe how to use the FP relation (Eq.\,\eqref{eq:FP_main}) to estimate the velocity dispersion $\sigma_{\rm FP}$, and in turn infer the boosted Einstein angle under the SIS assumption with $\sigma_v = \sigma_{\rm FP}$. To this end, it is necessary to understand how the effective radius $R_e$ and the mean surface brightness $I_e$ transform under a Lorentz boost. The radius can be estimated from the subtended angle and the lens redshift via a cosmological model. The subtended angle is subject to aberration, and the angular diameter distance $d\e{A}$ is estimated from an observed redshift, which is Doppler shifted by $\bs{v}_o$. The subtended angle $\alpha$ is affected on average as the Einstein angle (Eq.\,\eqref{eq:Einstein_Angle})
\begin{align}
\alpha = \alpha' \l( 1+ \frac{v_o}{c }\cos(\theta\e{cm}')\r)\,.
\end{align}
The radius can be expressed as
\begin{align}
R_e = \alpha d\e{A} = \alpha' d[z_l'] \l (1+ \frac{v_o}{c } \cos(\theta\e{cm}') + \frac{D_L}{d[z_l']} \frac{v_o}{c}\r)\,.
\end{align}
In the following, we show that the mean surface brightness is invariant under a boost. It is defined as (Eq.\,(1.20) p.23 of \cite{Binney:2008})
\begin{align}
I_e \equiv \frac{L}{2 \pi R_e^2} \,,
\end{align}
and is often quoted in [W pc$^{-2}$]. The luminosity $L$ [W] can be computed from the flux $L = 4 \pi d\e{L}^2 F$ using the luminosity distance $d\e{L}$ [m] and the flux $F$ [W m$^{-2}$], which can itself be computed from the specific intensity $I(\nu,\bs{\hat{n}})$ [W m$^{-2}$ Hz$^{-1}$ sr$^{-1}$] by integrating over frequency and solid angle over the relevant surface $\mathcal{S}$
\begin{align}
F = \int_{\mathbb{R}_+} \dd \nu  \int_{\mathcal{S}} \dd^2 \bs{\hat{n}} \, I(\nu,\bs{\hat{n}})R(\nu)\,.
\end{align}
where $R(\nu)$ is a filter that can be taken to be a tophat. Combining these steps, one obtains
\begin{align}
I_e = \frac{L}{2 \pi R_e^2} = \frac{4\pi d\e{L}^2 F}{2 d\e{A}^2 \Omega} = \frac{2 \pi  d\e{L}^2}{d\e{A}^2 \Omega} \int_{\mathbb{R}_+} \dd \nu \int_{\mathcal{S}}  \dd^2\bs{\hat{n}} \, I(\nu,\bs{\hat{n}}) R(\nu)
\end{align}
where $\Omega = \pi \alpha^2$. We assume for simplicity that $I_\nu(\nu,\bs{\hat{n}})$ is constant across the relevant angles spanned by $\hat{\bs{n}} \in \mathcal{S}$, i.e., $I(\nu,\bs{\hat{n}}) = I(\nu)$. This leads to 
\begin{align}
I_e = \frac{2 \pi  d\e{L}^2}{d\e{A}^2} \int_{\mathbb{R}_+} \dd \nu  \, I(\nu) R(\nu)\,. 
\end{align}
Using Etherington's reciprocity law $d\e{L} = (1+z_l)^2 d\e{A}$ \cite{Etherington:1933asu}, we obtain the simple expression
\begin{align}
I_e = 2\pi (1+z_l)^4 \int_{\mathbb{R}_+} \dd \nu I(\nu) R(\nu) \,.
\end{align}
One can use the following transformation properties
\begin{align}
I(\nu,\bs{\hat{n}}) & = I'(\nu',\bs{\hat{n}'}) \l( \frac{\nu}{\nu'}\r)^3\,,\\
(1+z_l) & = (1+z_l')\l(1 + Z_L \frac{v_o}{c}\r)\,, \\
\nu & = \frac{\nu}{\nu_s}  \frac{\nu_s}{\nu'} \nu' = \frac{(1+z_l')}{(1+z_l)} \nu'=\nu'  \l(1-Z_L \frac{v_o}{c}\r) \,,
\end{align}
where $Z_L$ was defined in Eq.\,\eqref{eq:Zl} in terms of the observer's lens and source peculiar velocities, and the identity for the specific intensity follows from the fact that $I/\nu^3$ is a Lorentz invariant. Combining these, the mean surface brightness transforms as
\begin{align}
I_e & = 2 \pi (1+z_l)^4 \int_{R_+} \dd \nu I(\nu) R(\nu)\\
& =  2 (1+z'_l)^4\l(1+4Z_L \frac{v_o}{c}\r)\int_{R_+} \dd \nu' \l(1-Z_L \frac{v_o}{c}\r) I'(\nu') \l( \frac{\nu}{\nu'}\r)^3 R(\nu') \nonumber \\
& = 2\pi (1+z'_l)^4 \l(1+ 4 Z_L \frac{v_o}{c}\r) \l(1- 4 Z_L \frac{v_o}{c}\r) \int_{R_+} \dd \nu' I'(\nu') R(\nu') \nonumber \\
& = 2\pi (1+z_l')^4 \int_{R_+} \dd \nu' I'(\nu') R(\nu') \nonumber  \\
& = I_e' \,.
\end{align}
Hence, the mean surface brightness is invariant under a boost. Hence, $\sigma_v$ is affected by the boost only via $R_e$. We obtain
\begin{align}
a \log_{10} \sigma_v & = \log_{10} R_e - b \log_{10} I_e - c \nonumber \\
& = \log_{10} (\alpha' d[z_l']) - b \log_{10}(I_e') - c  + \log_{10} \l (1+ \bs{v}_o \cdot \bs{\hat{n}'} + \frac{D_L}{d[z_l']} \frac{v_o}{c}\r) \,.
\end{align}
Explicitly, the velocity dispersion $\sigma_v$ can be expressed as
\begin{align}
\sigma_v = \sigma_v' \l( 1+ \Sigma_v \frac{v_o}{c}\r)\,,
\end{align}
with the naive $\sigma_v'$, which would be extracted from observables if we didn't account for the peculiar velocities and the correction $\Sigma_v$ defined as
\begin{align}
\sigma_v' & = \l( 10^{-c} \alpha' d[z_l'] (I_e')^{-b}\r)^{1/a} \,,\\
\Sigma_v & = \l(\frac{1}{a} \cos(\theta\e{cm}') +\frac{D_L}{a d[z_l']} \r) \,.
\end{align}
We can further compute the effect on the Einstein angle if one assumes that $\sigma_v = \sigma\e{FP}$ and adopts the SIS assumption
\begin{align}
\theta'_{E, \rm FP} = \frac{4\pi \sigma_v^2}{c^2} \frac{d_{l s}}{d_s} = \frac{4 \pi \sigma_v'^2}{c^2} \frac{d[z_l',z_s']}{d[z_s']} \l(1 + \frac{D_{LS}}{d[z_l',z_s']} \frac{v_o}{c} - \frac{D_{S}}{d[z_s']} \frac{v_o}{c}  + 2 \Sigma_v \frac{v_o}{c} \r) 
\end{align}
Replacing this expression for the Einstein angle in Eq.\,\eqref{eq:Einstein_Angle}, we get
\begin{align}
\theta'_{E, \rm FP} = \frac{4 \pi \sigma_v'^2 }{c^2} \frac{d[z_l',z_s']}{d[z_s']} \l(1 - \frac{v_o}{c}\cos(\theta'\e{cm})+ \frac{D_{LS}}{d[z_l',z_s']} \frac{v_o}{c} - \frac{D_{S}}{d[z_s']} \frac{v_o}{c}  + 2 \Sigma_v \frac{v_o}{c} \r) \,.
\end{align}
This expression can be used to infer $v_o$ from the observables $z_l'$, $z_s'$, $\alpha'$, $I_e'$, $\theta\e{cm}'$, and the FP parameters $a,b,c$, which we assume are not affected by the boost. This is the expression that appears in \eqref{eq:FP_Einstein_angle}.

\bibliographystyle{JHEP.bst}
\bibliography{references.bib}

\end{document}

%% file: packages.tex
\usepackage{jcappub}
\usepackage{siunitx}
\usepackage{mathrsfs}
\usepackage{bbm}
\usepackage{bm}
\usepackage{dsfont}
\usepackage{tensor}
\usepackage{xspace}
\usepackage{lmodern}
\usepackage{wasysym}
\usepackage{gensymb}
\usepackage{microtype}
\usepackage{empheq}
\usepackage{ifthen}
\usepackage{amsmath}
\usepackage{wrapfig}
\usepackage{cleveref}
\usepackage{xcolor}
\usepackage{siunitx}
\DeclareSIUnit\parsec{pc}
\usepackage{floatrow}
\usepackage[normalem]{ulem}
\usepackage{booktabs}
\usepackage{makecell}

%% file: macros.tex
\def \be {\begin{equation}}
\def \ee {\end{equation}}
\def \dd {\mathrm{d}}

\def \l {\left}
\def \r {\right}

\def \bs {\boldsymbol}

\newcommand{\e}[1]{_{\rm #1}}

\newcommand{\beq}{\begin{equation}}
\newcommand{\eeq}{\end{equation}}
\newcommand{\bea}{\begin{eqnarray}}
\newcommand{\eea}{\end{eqnarray}}

\newcommand{\thint}{\theta^{\rm int}}


\newcommand{\bn}{{\bf{n}}}

\newcommand{\HH}{{\cal H}}

\renewcommand\[{\left[}

\newcommand\ees{\end{eqnarray}}
\newcommand\bees{\begin{eqnarray}}
\def \bn {\boldsymbol{\hat{n}}}
\def\dd{\mathrm{d}}

\def \bs {\boldsymbol}
\def \be {\begin{equation}}
\def \ee {\end{equation}}

\renewcommand{\[}{\left[}

\def \dd {\mathrm{d}}

\def\arcsec{\hbox{$^{\prime\prime}$}}
\def\ksmpc{${\, \mathrm{km}\, \mathrm{s}^{-1}\, \mathrm{Mpc}^{-1}}$\xspace}
\def\ks{${\, \mathrm{km}\, \mathrm{s}^{-1}}$\xspace}


%% file: references.bib
@article{Cusin:2024git,
    author = "Cusin, Giulia and Pitrou, Cyril and Bonvin, Camille and Barrau, Aur{\'e}lien and Martineau, Killian",
    title = "{Boosting gravitational waves: a review of kinematic effects on amplitude, polarization, frequency and energy density}",
    eprint = "2405.01297",
    archivePrefix = "arXiv",
    primaryClass = "gr-qc",
    doi = "10.1088/1361-6382/ad7ad0",
    journal = "Class. Quant. Grav.",
    volume = "41",
    number = "22",
    pages = "225006",
    year = "2024"
}

@article{Dalang:2021ruy,
    author = "Dalang, Charles and Bonvin, Camille",
    title = "{On the kinematic cosmic dipole tension}",
    eprint = "2111.03616",
    archivePrefix = "arXiv",
    primaryClass = "astro-ph.CO",
    doi = "10.1093/mnras/stac726",
    journal = "Mon. Not. Roy. Astron. Soc.",
    volume = "512",
    number = "3",
    pages = "3895--3905",
    year = "2022"
}

@article{Schwarz:2016,
	doi = {10.1088/0264-9381/33/18/184001},
	url = {https://doi.org/10.1088/0264-9381/33/18/184001},
	year = 2016,
	month = {aug},
	publisher = {{IOP} Publishing},
	volume = {33},
	number = {18},
	pages = {184001},
	author = {Dominik J Schwarz and Craig J Copi and Dragan Huterer and Glenn D Starkman},
	title = {{CMB} anomalies after Planck},
	journal = {Classical and Quantum Gravity}
}

@article{Fixsen:1996,
    author = "Fixsen, D. J. and Cheng, E. S. and Gales, J. M. and Mather, John C. and Shafer, R. A. and Wright, E. L.",
    title = "{The Cosmic Microwave Background spectrum from the full COBE FIRAS data set}",
    eprint = "astro-ph/9605054",
    archivePrefix = "arXiv",
    doi = "10.1086/178173",
    journal = "Astrophys. J.",
    volume = "473",
    pages = "576",
    year = "1996"
}

@ARTICLE{Fixsen:1994,
       author = {{Fixsen}, D.~J. and {Cheng}, E.~S. and {Cottingham}, D.~A. and {Eplee}, R.~E., Jr. and {Isaacman}, R.~B. and {Mather}, J.~C. and {Meyer}, S.~S. and {Noerdlinger}, P.~D. and {Shafer}, R.~A. and {Weiss}, R. and {Wright}, E.~L. and {Bennett}, C.~L. and {Boggess}, N.~W. and {Kelsall}, T. and {Moseley}, S.~H. and {Silverberg}, R.~F. and {Smoot}, G.~F. and {Wilkinson}, D.~T.},
        title = "{Cosmic Microwave Background Dipole Spectrum Measured by the COBE FIRAS Instrument}",
      journal = {\apj},
         year = 1994,
        month = jan,
       volume = {420},
        pages = {445},
          doi = {10.1086/173575},
       adsurl = {https://ui.adsabs.harvard.edu/abs/1994ApJ...420..445F}
}

@article{Aghanim:2018eyx,
    author = "Aghanim, N. and others",
    collaboration = "Planck",
    title = "{Planck 2018 results. VI. Cosmological parameters}",
    eprint = "1807.06209",
    archivePrefix = "arXiv",
    primaryClass = "astro-ph.CO",
    doi = "10.1051/0004-6361/201833910",
    journal = "Astron. Astrophys.",
    volume = "641",
    pages = "A6",
    year = "2020"
}

@article{Planck:2013kqc,
    author = "Aghanim, N. and others",
    collaboration = "Planck",
    title = "{Planck 2013 results. XXVII. Doppler boosting of the CMB: Eppur si muove}",
    eprint = "1303.5087",
    archivePrefix = "arXiv",
    primaryClass = "astro-ph.CO",
    doi = "10.1051/0004-6361/201321556",
    journal = "Astron. Astrophys.",
    volume = "571",
    pages = "A27",
    year = "2014"
}

@article{Saha:2021bay,
    author = "Saha, Sayan and Shaikh, Shabbir and Mukherjee, Suvodip and Souradeep, Tarun and Wandelt, Benjamin D.",
    title = "{Bayesian estimation of our local motion from the Planck-2018 CMB temperature map}",
    eprint = "2106.07666",
    archivePrefix = "arXiv",
    primaryClass = "astro-ph.CO",
    doi = "10.1088/1475-7516/2021/10/072",
    journal = "JCAP",
    volume = "10",
    pages = "072",
    year = "2021"
}

@article{Ferreira:2020aqa,
    author = "Ferreira, Pedro da Silveira and Quartin, Miguel",
    title = "{First Constraints on the Intrinsic CMB Dipole and Our Velocity with Doppler and Aberration}",
    eprint = "2011.08385",
    archivePrefix = "arXiv",
    primaryClass = "astro-ph.CO",
    doi = "10.1103/PhysRevLett.127.101301",
    journal = "Phys. Rev. Lett.",
    volume = "127",
    number = "10",
    pages = "101301",
    year = "2021"
}

@ARTICLE{Ellis1984,
       author = {{Ellis}, G.~F.~R. and {Baldwin}, J.~E.},
        title = "{On the expected anisotropy of radio source counts}",
      journal = {MNRAS},
         year = 1984,
        month = jan,
       volume = {206},
        pages = {377-381},
          doi = {10.1093/mnras/206.2.377},
       adsurl = {https://ui.adsabs.harvard.edu/abs/1984MNRAS.206..377E}
}

@article{Secrest:2020has,
   title={A Test of the Cosmological Principle with Quasars},
   volume={908},
   ISSN={2041-8213},
   url={http://dx.doi.org/10.3847/2041-8213/abdd40},
   DOI={10.3847/2041-8213/abdd40},
   number={2},
   journal={The Astrophysical Journal},
   publisher={American Astronomical Society},
   author={Secrest, Nathan J. and Hausegger, Sebastian von and Rameez, Mohamed and Mohayaee, Roya and Sarkar, Subir and Colin, Jacques},
   year={2021},
   month={Feb},
   pages={L51}
}

@article{Secrest:2022uvx,
    author = "Secrest, Nathan and von Hausegger, Sebastian and Rameez, Mohamed and Mohayaee, Roya and Sarkar, Subir",
    title = "{A Challenge to the Standard Cosmological Model}",
    eprint = "2206.05624",
    archivePrefix = "arXiv",
    primaryClass = "astro-ph.CO",
    month = "6",
    year = "2022"
}

@ARTICLE{Said2025,
       author = {{Said}, Khaled and {Howlett}, Cullan and {Davis}, Tamara and {Lucey}, John and {Saulder}, Christoph and {Douglass}, Kelly and {Kim}, Alex G. and {Kremin}, Anthony and {Ross}, Caitlin and {Aldering}, Greg and {Aguilar}, Jessica Nicole and {Ahlen}, Steven and {BenZvi}, Segev and {Bianchi}, Davide and {Brooks}, David and {Claybaugh}, Todd and {Dawson}, Kyle and {de la Macorra}, Axel and {Dey}, Biprateep and {Doel}, Peter and {Fanning}, Kevin and {Ferraro}, Simone and {Font-Ribera}, Andreu and {Forero-Romero}, Jaime E. and {Gazta{\~n}aga}, Enrique and {Gontcho}, Satya Gontcho A. and {Guy}, Julien and {Honscheid}, Klaus and {Kehoe}, Robert and {Kisner}, Theodore and {Lambert}, Andrew and {Landriau}, Martin and {Le Guillou}, Laurent and {Manera}, Marc and {Meisner}, Aaron and {Miquel}, Ramon and {Moustakas}, John and {Mu{\~n}oz-Guti{\'e}rrez}, Andrea and {Myers}, Adam and {Nie}, Jundan and {Palanque-Delabrouille}, Nathalie and {Percival}, Will and {Prada}, Francisco and {Rossi}, Graziano and {Sanchez}, Eusebio and {Schlegel}, David and {Schubnell}, Michael and {Silber}, Joseph Harry and {Sprayberry}, David and {Tarl{\'e}}, Gregory and {Magana}, Mariana Vargas and {Weaver}, Benjamin Alan and {Wechsler}, Risa and {Zhou}, Zhimin and {Zou}, Hu},
        title = "{DESI peculiar velocity survey {\textendash} Fundamental Plane}",
      journal = {\mnras},
         year = 2025,
        month = jun,
       volume = {539},
       number = {4},
        pages = {3627-3644},
          doi = {10.1093/mnras/staf700},
archivePrefix = {arXiv},
       eprint = {2408.13842},
 primaryClass = {astro-ph.CO},
       adsurl = {https://ui.adsabs.harvard.edu/abs/2025MNRAS.539.3627S}
}

@article{Bonvin:2005ps,
    author = "Bonvin, Camille and Durrer, Ruth and Gasparini, M. Alice",
    title = "{Fluctuations of the luminosity distance}",
    eprint = "astro-ph/0511183",
    archivePrefix = "arXiv",
    doi = "10.1103/PhysRevD.85.029901",
    journal = "Phys. Rev. D",
    volume = "73",
    pages = "023523",
    year = "2006",
    note = "[Erratum: Phys.Rev.D 85, 029901 (2012)]"
}

@article{Dam:2022wwh,
    author = "Dam, Lawrence and Lewis, Geraint F. and Brewer, Brendon J.",
    title = "{Testing the Cosmological Principle with CatWISE Quasars: A Bayesian Analysis of the Number-Count Dipole}",
    eprint = "2212.07733",
    archivePrefix = "arXiv",
    primaryClass = "astro-ph.CO",
    month = "12",
    year = "2022"
}

@ARTICLE{Sereno2008,
       author = {{Sereno}, M.},
        title = "{Aberration in gravitational lensing}",
      journal = {\prd},
         year = 2008,
        month = oct,
       volume = {78},
       number = {8},
          eid = {083003},
        pages = {083003},
          doi = {10.1103/PhysRevD.78.083003},
archivePrefix = {arXiv},
       eprint = {0809.3900},
 primaryClass = {astro-ph},
       adsurl = {https://ui.adsabs.harvard.edu/abs/2008PhRvD..78h3003S}
}

@ARTICLE{Dalang2023,
       author = {{Dalang}, Charles and {Millon}, Martin and {Baker}, Tessa},
        title = "{Peculiar velocity effects on the Hubble constant from time-delay cosmography}",
      journal = {\prd},
         year = 2023,
        month = jun,
       volume = {107},
       number = {12},
          eid = {123528},
        pages = {123528},
          doi = {10.1103/PhysRevD.107.123528},
archivePrefix = {arXiv},
       eprint = {2301.05574},
 primaryClass = {astro-ph.CO},
       adsurl = {https://ui.adsabs.harvard.edu/abs/2023PhRvD.107l3528D}
}

@ARTICLE{Djorgovski1987,
       author = {{Djorgovski}, S. and {Davis}, Marc},
        title = "{Fundamental Properties of Elliptical Galaxies}",
      journal = {\apj},
         year = 1987,
        month = feb,
       volume = {313},
        pages = {59},
          doi = {10.1086/164948},
       adsurl = {https://ui.adsabs.harvard.edu/abs/1987ApJ...313...59D}
}

@ARTICLE{Dressler1987,
       author = {{Dressler}, Alan and {Lynden-Bell}, Donald and {Burstein}, David and {Davies}, Roger L. and {Faber}, S.~M. and {Terlevich}, Roberto and {Wegner}, Gary},
        title = "{Spectroscopy and Photometry of Elliptical Galaxies. I. New Distance Estimator}",
      journal = {\apj},
         year = 1987,
        month = feb,
       volume = {313},
        pages = {42},
          doi = {10.1086/164947},
       adsurl = {https://ui.adsabs.harvard.edu/abs/1987ApJ...313...42D}
}

@ARTICLE{Mittal2024,
       author = {{Mittal}, Vasudev and {Oayda}, Oliver T. and {Lewis}, Geraint F.},
        title = "{The cosmic dipole in the Quaia sample of quasars: a Bayesian analysis}",
      journal = {\mnras},
         year = 2024,
        month = jan,
       volume = {527},
       number = {3},
        pages = {8497-8510},
          doi = {10.1093/mnras/stad3706},
archivePrefix = {arXiv},
       eprint = {2311.14938},
 primaryClass = {astro-ph.CO},
       adsurl = {https://ui.adsabs.harvard.edu/abs/2024MNRAS.527.8497M}
}

@ARTICLE{Abghari2024,
       author = {{Abghari}, Arefe and {Bunn}, Emory F. and {Hergt}, Lukas T. and {Li}, Boris and {Scott}, Douglas and {Sullivan}, Raelyn M. and {Wei}, Dingchen},
        title = "{Reassessment of the dipole in the distribution of quasars on the sky}",
      journal = {\jcap},
         year = 2024,
        month = nov,
       volume = {2024},
       number = {11},
          eid = {067},
        pages = {067},
          doi = {10.1088/1475-7516/2024/11/067},
archivePrefix = {arXiv},
       eprint = {2405.09762},
 primaryClass = {astro-ph.CO},
       adsurl = {https://ui.adsabs.harvard.edu/abs/2024JCAP...11..067A}
}

@ARTICLE{Oayda2024,
       author = {{Oayda}, Oliver T. and {Mittal}, Vasudev and {Lewis}, Geraint F. and {Murphy}, Tara},
        title = "{A Bayesian approach to the cosmic dipole in radio galaxy surveys: joint analysis of NVSS \& RACS}",
      journal = {\mnras},
         year = 2024,
        month = jul,
       volume = {531},
       number = {4},
        pages = {4545-4559},
          doi = {10.1093/mnras/stae1399},
archivePrefix = {arXiv},
       eprint = {2406.01871},
 primaryClass = {astro-ph.CO},
       adsurl = {https://ui.adsabs.harvard.edu/abs/2024MNRAS.531.4545O}
}

@ARTICLE{Bashir2025,
       author = {{Bashir}, Masroor and {Chingangbam}, Pravabati and {Appleby}, Stephen},
        title = "{The CatWISE2020 Quasar dipole: A Reassessment of the Cosmic Dipole Anomaly}",
      journal = {arXiv e-prints},
         year = 2025,
        month = nov,
          eid = {arXiv:2511.00822},
        pages = {arXiv:2511.00822},
          doi = {10.48550/arXiv.2511.00822},
archivePrefix = {arXiv},
       eprint = {2511.00822},
 primaryClass = {astro-ph.CO},
       adsurl = {https://ui.adsabs.harvard.edu/abs/2025arXiv251100822B}
}

@misc{LensPop,
       author = {{Collett}, Thomas E.},
        title = "{LensPop: Galaxy-galaxy strong lensing population simulation}",
 howpublished = {Astrophysics Source Code Library, record ascl:1705.009},
         year = 2017,
        month = may,
          eid = {ascl:1705.009},
archivePrefix = {ascl},
       eprint = {1705.009},
       adsurl = {https://ui.adsabs.harvard.edu/abs/2017ascl.soft05009C}
}

@ARTICLE{Said2020,
       author = {{Said}, Khaled and {Colless}, Matthew and {Magoulas}, Christina and {Lucey}, John R. and {Hudson}, Michael J.},
        title = "{Joint analysis of 6dFGS and SDSS peculiar velocities for the growth rate of cosmic structure and tests of gravity}",
      journal = {\mnras},
         year = 2020,
        month = sep,
       volume = {497},
       number = {1},
        pages = {1275-1293},
          doi = {10.1093/mnras/staa2032},
archivePrefix = {arXiv},
       eprint = {2007.04993},
 primaryClass = {astro-ph.CO},
       adsurl = {https://ui.adsabs.harvard.edu/abs/2020MNRAS.497.1275S}
}

@ARTICLE{Magoulas2012,
       author = {{Magoulas}, Christina and {Springob}, Christopher M. and {Colless}, Matthew and {Jones}, D. Heath and {Campbell}, Lachlan A. and {Lucey}, John R. and {Mould}, Jeremy and {Jarrett}, Tom and {Merson}, Alex and {Brough}, Sarah},
        title = "{The 6dF Galaxy Survey: the near-infrared Fundamental Plane of early-type galaxies}",
      journal = {\mnras},
         year = 2012,
        month = nov,
       volume = {427},
       number = {1},
        pages = {245-273},
          doi = {10.1111/j.1365-2966.2012.21421.x},
archivePrefix = {arXiv},
       eprint = {1206.0385},
 primaryClass = {astro-ph.CO},
       adsurl = {https://ui.adsabs.harvard.edu/abs/2012MNRAS.427..245M}
}

@ARTICLE{Colless2001b,
       author = {{Colless}, Matthew and {Saglia}, R.~P. and {Burstein}, David and {Davies}, Roger L. and {McMahan}, Robert K. and {Wegner}, Gary},
        title = "{The peculiar motions of early-type galaxies in two distant regions - VII. Peculiar velocities and bulk motions}",
      journal = {\mnras},
         year = 2001,
        month = feb,
       volume = {321},
       number = {2},
        pages = {277-305},
          doi = {10.1046/j.1365-8711.2001.04044.x},
archivePrefix = {arXiv},
       eprint = {astro-ph/0008418},
 primaryClass = {astro-ph},
       adsurl = {https://ui.adsabs.harvard.edu/abs/2001MNRAS.321..277C}
}

@ARTICLE{Storey-Fisher2024,
       author = {{Storey-Fisher}, Kate and {Hogg}, David W. and {Rix}, Hans-Walter and {Eilers}, Anna-Christina and {Fabbian}, Giulio and {Blanton}, Michael R. and {Alonso}, David},
        title = "{Quaia, the Gaia-unWISE Quasar Catalog: An All-sky Spectroscopic Quasar Sample}",
      journal = {\apj},
         year = 2024,
        month = mar,
       volume = {964},
       number = {1},
          eid = {69},
        pages = {69},
          doi = {10.3847/1538-4357/ad1328},
archivePrefix = {arXiv},
       eprint = {2306.17749},
 primaryClass = {astro-ph.GA},
       adsurl = {https://ui.adsabs.harvard.edu/abs/2024ApJ...964...69S}
}

@ARTICLE{secrest2025,
       author = {{Secrest}, Nathan J.},
        title = "{The Ellis-Baldwin test}",
      journal = {Philosophical Transactions of the Royal Society of London Series A},
         year = 2025,
        month = feb,
       volume = {383},
       number = {2290},
          eid = {20240027},
        pages = {20240027},
          doi = {10.1098/rsta.2024.0027},
       adsurl = {https://ui.adsabs.harvard.edu/abs/2025RSPTA.38340027S}
}

@ARTICLE{Secrest2025c,
       author = {{Secrest}, Nathan and {von Hausegger}, Sebastian and {Rameez}, Mohamed and {Mohayaee}, Roya and {Sarkar}, Subir},
        title = "{Colloquium: The Cosmic Dipole Anomaly}",
      journal = {arXiv e-prints},
         year = 2025,
        month = may,
          eid = {arXiv:2505.23526},
        pages = {arXiv:2505.23526},
          doi = {10.48550/arXiv.2505.23526},
archivePrefix = {arXiv},
       eprint = {2505.23526},
 primaryClass = {astro-ph.CO},
       adsurl = {https://ui.adsabs.harvard.edu/abs/2025arXiv250523526S}
}

@ARTICLE{EuclidHolloway2025,
       author = {{Holloway}, P. and others},
        title = "{Euclid Quick Data Release (Q1). The Strong Lensing Discovery Engine E -- Ensemble classification of strong gravitational lenses: lessons for Data Release 1}",
      collaboration = "Euclid",
      journal = {arXiv e-prints},
         year = 2025,
        month = mar,
          eid = {arXiv:2503.15328},
        pages = {arXiv:2503.15328},
          doi = {10.48550/arXiv.2503.15328},
archivePrefix = {arXiv},
       eprint = {2503.15328},
 primaryClass = {astro-ph.GA},
       adsurl = {https://ui.adsabs.harvard.edu/abs/2025arXiv250315328E}
}

@ARTICLE{EuclidWalmsley2025,
       author = {{Walmsley}, M. and others},
       collaboration = "Euclid",
        title = "{Euclid Quick Data Release (Q1): The Strong Lensing Discovery Engine A -- System overview and lens catalogue}",
      journal = {arXiv e-prints},
         year = 2025,
        month = mar,
          eid = {arXiv:2503.15324},
        pages = {arXiv:2503.15324},
          doi = {10.48550/arXiv.2503.15324},
archivePrefix = {arXiv},
       eprint = {2503.15324},
 primaryClass = {astro-ph.GA},
       adsurl = {https://ui.adsabs.harvard.edu/abs/2025arXiv250315324E}
}

@ARTICLE{Euclid2025Rojas,
       author = {{Rojas}, K. and others},
        collaboration = "Euclid",
        title = "{Euclid Quick Data Release (Q1) The Strong Lensing Discovery Engine B -- Early strong lens candidates from visual inspection of high velocity dispersion galaxies}",
      journal = {arXiv e-prints},
         year = 2025,
        month = mar,
          eid = {arXiv:2503.15325},
        pages = {arXiv:2503.15325},
          doi = {10.48550/arXiv.2503.15325},
archivePrefix = {arXiv},
       eprint = {2503.15325},
 primaryClass = {astro-ph.GA},
       adsurl = {https://ui.adsabs.harvard.edu/abs/2025arXiv250315325E}
}

@ARTICLE{EuclidLines2025,
       author = {{Lines}, N.~E.~P. and others},
    collaboration = "Euclid",
        title = "{Euclid Quick Data Release (Q1). The Strong Lensing Discovery Engine C: Finding lenses with machine learning}",
      journal = {arXiv e-prints},
         year = 2025,
        month = mar,
          eid = {arXiv:2503.15326},
        pages = {arXiv:2503.15326},
          doi = {10.48550/arXiv.2503.15326},
archivePrefix = {arXiv},
       eprint = {2503.15326},
 primaryClass = {astro-ph.GA},
       adsurl = {https://ui.adsabs.harvard.edu/abs/2025arXiv250315326E}
}

@BOOK{Binney:2008,
       author = {{Binney}, James and {Tremaine}, Scott},
        title = "{Galactic Dynamics: Second Edition}",
         year = 2008,
       adsurl = {https://ui.adsabs.harvard.edu/abs/2008gady.book.....B}
}

@article{Etherington:1933asu,
    author = "Etherington, I. M. H.",
    title = "{LX. On the definition of distance in general relativity}",
    doi = "10.1080/14786443309462220",
    month = "4",
    year = "1933"
}

@techreport{EuclidSciRD2010,
  author      = {Laureijs, R. and others},
  title       = {{Euclid Science Requirements Document}},
  institution = {European Space Agency},
  year        = {2010},
  type        = {ESA Science Document},
  number      = {DEM-SA-Dc-00001},
  issue       = {4},
  revision    = {0},
  month       = mar,
  url         = {https://sci.esa.int/documents/33220/36137/1567257215944-Euclid_SciRD_DEM-SA-DC-0001_4_0_2010-03-22.pdf},
  note        = {Issue 4.0, 5 March 2010}
}

@ARTICLE{EuclidBretonniere2023,
       author = {{Bretonni{\`e}re}, H. and others},
        title = "{Euclid preparation. XXVI. The Euclid Morphology Challenge: Towards structural parameters for billions of galaxies}",
     collaboration = "Euclid",
      journal = {\aap},
         year = 2023,
        month = mar,
       volume = {671},
          eid = {A102},
        pages = {A102},
          doi = {10.1051/0004-6361/202245042},
archivePrefix = {arXiv},
       eprint = {2209.12907},
 primaryClass = {astro-ph.GA},
       adsurl = {https://ui.adsabs.harvard.edu/abs/2023A&A...671A.102E}
}

@ARTICLE{Collett2016,
       author = {{Collett}, Thomas E. and {Cunnington}, Steven D.},
        title = "{Observational selection biases in time-delay strong lensing and their impact on cosmography}",
      journal = {\mnras},
         year = 2016,
        month = nov,
       volume = {462},
       number = {3},
        pages = {3255-3264},
          doi = {10.1093/mnras/stw1856},
archivePrefix = {arXiv},
       eprint = {1605.08341},
 primaryClass = {astro-ph.CO},
       adsurl = {https://ui.adsabs.harvard.edu/abs/2016MNRAS.462.3255C}
}

@ARTICLE{Collett2015,
       author = {{Collett}, Thomas E.},
        title = "{The Population of Galaxy-Galaxy Strong Lenses in Forthcoming Optical Imaging Surveys}",
      journal = {\apj},
         year = 2015,
        month = sep,
       volume = {811},
       number = {1},
          eid = {20},
        pages = {20},
          doi = {10.1088/0004-637X/811/1/20},
archivePrefix = {arXiv},
       eprint = {1507.02657},
 primaryClass = {astro-ph.CO},
       adsurl = {https://ui.adsabs.harvard.edu/abs/2015ApJ...811...20C}
}

@INPROCEEDINGS{Connolly2010,
       author = {{Connolly}, A.~J. and {Peterson}, John and {Jernigan}, J. Garrett and {Abel}, Robert and {Bankert}, Justin and {Chang}, Chihway and {Claver}, Charles F. and {Gibson}, Robert and {Gilmore}, David K. and {Grace}, Emily and {Jones}, R. Lynne and {Ivezic}, Zeljko and {Jee}, James and {Juric}, Mario and {Kahn}, Steven M. and {Krabbendam}, Victor L. and {Krughoff}, Simon and {Lorenz}, Suzanne and {Pizagno}, James and {Rasmussen}, Andrew and {Todd}, Nathan and {Tyson}, J. Anthony and {Young}, Mallory},
        title = "{Simulating the LSST system}",
    booktitle = {Modeling, Systems Engineering, and Project Management for Astronomy IV},
         year = 2010,
       editor = {{Angeli}, George Z. and {Dierickx}, Philippe},
       series = {Society of Photo-Optical Instrumentation Engineers (SPIE) Conference Series},
       volume = {7738},
        month = jul,
          eid = {77381O},
        pages = {77381O},
          doi = {10.1117/12.857819},
       adsurl = {https://ui.adsabs.harvard.edu/abs/2010SPIE.7738E..1OC}
}

@ARTICLE{DeLucia2006,
       author = {{De Lucia}, Gabriella and {Springel}, Volker and {White}, Simon D.~M. and {Croton}, Darren and {Kauffmann}, Guinevere},
        title = "{The formation history of elliptical galaxies}",
      journal = {\mnras},
         year = 2006,
        month = feb,
       volume = {366},
       number = {2},
        pages = {499-509},
          doi = {10.1111/j.1365-2966.2005.09879.x},
archivePrefix = {arXiv},
       eprint = {astro-ph/0509725},
 primaryClass = {astro-ph},
       adsurl = {https://ui.adsabs.harvard.edu/abs/2006MNRAS.366..499D}
}

@ARTICLE{Sonnenfeld2023,
       author = {{Sonnenfeld}, Alessandro and {Li}, Shun-Sheng and {Despali}, Giulia and {Gavazzi}, Raphael and {Shajib}, Anowar J. and {Taylor}, Edward N.},
        title = "{Strong lensing selection effects}",
      journal = {\aap},
         year = 2023,
        month = oct,
       volume = {678},
          eid = {A4},
        pages = {A4},
          doi = {10.1051/0004-6361/202346026},
archivePrefix = {arXiv},
       eprint = {2301.13230},
 primaryClass = {astro-ph.GA},
       adsurl = {https://ui.adsabs.harvard.edu/abs/2023A&A...678A...4S}
}

@ARTICLE{Sonnenfeld2022,
       author = {{Sonnenfeld}, Alessandro},
        title = "{Statistical strong lensing. III. Inferences with complete samples of lenses}",
      journal = {\aap},
         year = 2022,
        month = mar,
       volume = {659},
          eid = {A132},
        pages = {A132},
          doi = {10.1051/0004-6361/202142301},
archivePrefix = {arXiv},
       eprint = {2109.13246},
 primaryClass = {astro-ph.GA},
       adsurl = {https://ui.adsabs.harvard.edu/abs/2022A&A...659A.132S}
}

@ARTICLE{Herle2024,
       author = {{Herle}, A. and {O'Riordan}, C.~M. and {Vegetti}, S.},
        title = "{Selection functions of strong lens finding neural networks}",
      journal = {\mnras},
         year = 2024,
        month = oct,
       volume = {534},
       number = {2},
        pages = {1093-1106},
          doi = {10.1093/mnras/stae2106},
archivePrefix = {arXiv},
       eprint = {2307.10355},
 primaryClass = {astro-ph.CO},
       adsurl = {https://ui.adsabs.harvard.edu/abs/2024MNRAS.534.1093H}
}

@article{nautilus,
    author = {Lange, Johannes U},
    title = "{nautilus: boosting Bayesian importance nested sampling with deep learning}",
    journal = {Monthly Notices of the Royal Astronomical Society},
    volume = {525},
    number = {2},
    pages = {3181-3194},
    year = {2023},
    month = {08},
    doi = {10.1093/mnras/stad2441},
    url = {https://doi.org/10.1093/mnras/stad2441},
    eprint = {https://academic.oup.com/mnras/article-pdf/525/2/3181/51331635/stad2441.pdf},
}

@ARTICLE{4MOSToverview,
       author = {{de Jong}, R.~S. and others},
        collaboration = "4MOST",
        title = "{4MOST: Project overview and information for the First Call for Proposals}",
      journal = {The Messenger},
         year = 2019,
        month = mar,
       volume = {175},
        pages = {3-11},
          doi = {10.18727/0722-6691/5117},
archivePrefix = {arXiv},
       eprint = {1903.02464},
 primaryClass = {astro-ph.IM},
       adsurl = {https://ui.adsabs.harvard.edu/abs/2019Msngr.175....3D}
}

@ARTICLE{DESI2016,
       author = {{DESI Collaboration} and {Aghamousa}, Amir and others},
        collaboration = "DESI",
        title = "{The DESI Experiment Part I: Science,Targeting, and Survey Design}",
      journal = {arXiv e-prints},
         year = 2016,
        month = oct,
          eid = {arXiv:1611.00036},
        pages = {arXiv:1611.00036},
          doi = {10.48550/arXiv.1611.00036},
archivePrefix = {arXiv},
       eprint = {1611.00036},
 primaryClass = {astro-ph.IM},
       adsurl = {https://ui.adsabs.harvard.edu/abs/2016arXiv161100036D}
}

@ARTICLE{4SLSLS,
       author = {{Collett}, T.~E. and {Sonnenfeld}, A. and {Frohmaier}, C. and {Glazebrook}, K. and {Sluse}, D. and {Motta}, V. and {Verma}, A. and {Anguita}, T. and {Koopmans}, L. and {Tortora}, C. and {Courbin}, F. and {Cabanac}, R. and {Frye}, B. and {Smith}, G.~P. and {Diego}, J.~M. and {Alteiri}, B. and {Lopez}, S. and {Fassnacht}, C. and {Cooray}, A. and {Goobar}, A. and {Ryczanowski}, D. and {Serjeant}, S. and {Richard}, J. and {Treu}, T. and {Moustakas}, L. and {Li}, R. and {Jacobs}, C. and {Lemon}, C. and {Marchetti}, L. and {Hartley}, P. and {Jullo}, E. and {Lee}, C.-H. and {Birrer}, S. and {Fritz}, A. and {Nightingale}, J. and {Napolitano}, N. and {Plazas}, A.~A. and {Kruk}, S. and {Spiniello}, C. and {Grillo}, C. and {Suyu}, S. and {Shajib}, A. and {Vernardos}, G. and {Dye}, S. and {Daylan}, T. and {Newman}, J. and {Schuldt}, S.},
        title = "{The 4MOST Strong Lensing Spectroscopic Legacy Survey (4SLSLS)}",
      journal = {The Messenger},
         year = 2023,
        month = mar,
       volume = {190},
        pages = {49-52},
          doi = {10.18727/0722-6691/5313},
       adsurl = {https://ui.adsabs.harvard.edu/abs/2023Msngr.190...49C}
}

@ARTICLE{Huang2025,
       author = {{Huang}, Xiaosheng and {Inchausti}, Jose Carlos and {Storfer}, Christopher J. and {Tabares-Tarquinio}, S. and {Moustakas}, J. and {Sheu}, W. and {Agarwal}, S. and {Tamargo-Arizmendi}, M. and {Schlegel}, D.~J. and {Aguilar}, J. and {Ahlen}, S. and {Aldering}, G. and {Bailey}, S. and {Banka}, S. and {BenZvi}, S. and {Bianchi}, D. and {Bolton}, A. and {Brooks}, D. and {Cikota}, A. and {Claybaugh}, T. and {Dawson}, K.~S. and {de la Macorra}, A. and {Dey}, A. and {Doel}, P. and {Edelstein}, J. and {Forero-Romero}, J.~E. and {Gaztanaga}, E. and {Gontcho}, S. Gontcho A and {Gonzalez-Morales}, A.~X. and {Gu}, A. and {Honscheid}, K. and {Ishak}, M. and {Juneau}, S. and {Kehoe}, R. and {Kisner}, T. and {Koposov}, S.~E. and {Kwon}, K.~J. and {Lambert}, A. and {Landriau}, M. and {Lang}, D. and {Le Guillou}, L. and {Levi}, M.~E. and {Liu}, J. and {Meisner}, A. and {Miquel}, R. and {Myers}, A.~D. and {Perlmutter}, S. and {Palanque-Delabrouille}, N. and {Perez-Rafols}, I. and {Poppett}, C. and {Prada}, F. and {Rossi}, G. and {Rubin}, D. and {Sanchez}, E. and {Schubnell}, M. and {Shu}, Y. and {Silver}, E. and {Sprayberry}, D. and {Suzuki}, N. and {Tarle}, G. and {Weaver}, B.~A. and {Zou}, H.},
        title = "{DESI Strong Lens Foundry II: DESI Spectroscopy for Strong Lens Candidates}",
      journal = {arXiv e-prints},
         year = 2025,
        month = sep,
          eid = {arXiv:2509.18089},
        pages = {arXiv:2509.18089},
          doi = {10.48550/arXiv.2509.18089},
archivePrefix = {arXiv},
       eprint = {2509.18089},
 primaryClass = {astro-ph.CO},
       adsurl = {https://ui.adsabs.harvard.edu/abs/2025arXiv250918089H}
}

@ARTICLE{Park2007,
       author = {{Park}, Changbom and {Choi}, Yun-Young and {Vogeley}, Michael S. and {Gott}, III, J. Richard and {Blanton}, Michael R. and {SDSS Collaboration}},
        title = "{Environmental Dependence of Properties of Galaxies in the Sloan Digital Sky Survey}",
      journal = {\apj},
         year = 2007,
        month = apr,
       volume = {658},
       number = {2},
        pages = {898-916},
          doi = {10.1086/511059},
archivePrefix = {arXiv},
       eprint = {astro-ph/0611610},
 primaryClass = {astro-ph},
       adsurl = {https://ui.adsabs.harvard.edu/abs/2007ApJ...658..898P}
}

@ARTICLE{AcevedoBarroso2025,
       author = {{Acevedo Barroso}, J.~A. and {O'Riordan}, C.~M. and {Cl{\'e}ment}, B. and {Tortora}, C. and {Collett}, T.~E. and {Courbin}, F. and {Gavazzi}, R. and {Metcalf}, R.~B. and {Busillo}, V. and {Andika}, I.~T. and {Cabanac}, R. and {Courtois}, H.~M. and {Crook-Mansour}, J. and {Delchambre}, L. and {Despali}, G. and {Ecker}, L.~R. and {Franco}, A. and {Holloway}, P. and {Jackson}, N. and {Jahnke}, K. and {Mahler}, G. and {Marchetti}, L. and {Matavulj}, P. and {Melo}, A. and {Meneghetti}, M. and {Moustakas}, L.~A. and {M{\"u}ller}, O. and {Nucita}, A.~A. and {Paulino-Afonso}, A. and {Pearson}, J. and {Rojas}, K. and {Scarlata}, C. and {Schuldt}, S. and {Serjeant}, S. and {Sluse}, D. and {Suyu}, S.~H. and {Vaccari}, M. and {Verma}, A. and {Vernardos}, G. and {Walmsley}, M. and {Bouy}, H. and {Walth}, G.~L. and {Powell}, D.~M. and {Bolzonella}, M. and {Cuillandre}, J.-C. and {Kluge}, M. and {Saifollahi}, T. and {Schirmer}, M. and {Stone}, C. and {Acebron}, A. and {Bazzanini}, L. and {D{\'\i}az-S{\'a}nchez}, A. and {Hogg}, N.~B. and {Koopmans}, L.~V.~E. and {Kruk}, S. and {Leuzzi}, L. and {Manj{\'o}n-Garc{\'\i}a}, A. and {Mannucci}, F. and {Nagam}, B.~C. and {Pearce-Casey}, R. and {Scharr{\'e}}, L. and {Wilde}, J. and {Altieri}, B. and {Amara}, A. and {Andreon}, S. and {Auricchio}, N. and {Baccigalupi}, C. and {Baldi}, M. and {Balestra}, A. and {Bardelli}, S. and {Basset}, A. and {Battaglia}, P. and {Bender}, R. and {Bonino}, D. and {Branchini}, E. and {Brescia}, M. and {Brinchmann}, J. and {Caillat}, A. and {Camera}, S. and {Candini}, G.~P. and {Capobianco}, V. and {Carbone}, C. and {Carretero}, J. and {Casas}, S. and {Castellano}, M. and {Castignani}, G. and {Cavuoti}, S. and {Cimatti}, A. and {Colodro-Conde}, C. and {Congedo}, G. and {Conselice}, C.~J. and {Conversi}, L. and {Copin}, Y. and {Corcione}, L. and {Cropper}, M. and {Da Silva}, A. and {Degaudenzi}, H. and {De Lucia}, G. and {Dinis}, J. and {Dubath}, F. and {Dupac}, X. and {Dusini}, S. and {Farina}, M. and {Farrens}, S. and {Ferriol}, S. and {Frailis}, M. and {Franceschi}, E. and {Galeotta}, S. and {Garilli}, B. and {George}, K. and {Gillard}, W. and {Gillis}, B. and {Giocoli}, C. and {G{\'o}mez-Alvarez}, P. and {Grazian}, A. and {Grupp}, F. and {Guzzo}, L. and {Haugan}, S.~V.~H. and {Hoekstra}, H. and {Holmes}, W. and {Hook}, I. and {Hormuth}, F. and {Hornstrup}, A. and {Jhabvala}, M. and {Joachimi}, B. and {Keih{\"a}nen}, E. and {Kermiche}, S. and {Kiessling}, A. and {Kubik}, B. and {Kunz}, M. and {Kurki-Suonio}, H. and {Le Mignant}, D. and {Ligori}, S. and {Lilje}, P.~B. and {Lindholm}, V. and {Lloro}, I. and {Mainetti}, G. and {Maiorano}, E. and {Mansutti}, O. and {Marcin}, S. and {Marggraf}, O. and {Martinelli}, M. and {Martinet}, N. and {Marulli}, F. and {Massey}, R. and {Medinaceli}, E. and {Melchior}, M. and {Mellier}, Y. and {Merlin}, E. and {Meylan}, G. and {Moresco}, M. and {Moscardini}, L. and {Munari}, E. and {Nakajima}, R. and {Neissner}, C. and {Nichol}, R.~C. and {Niemi}, S.-M. and {Nightingale}, J.~W. and {Padilla}, C. and {Paltani}, S. and {Pasian}, F. and {Pedersen}, K. and {Percival}, W.~J. and {Pettorino}, V. and {Pires}, S. and {Polenta}, G. and {Poncet}, M. and {Popa}, L.~A. and {Pozzetti}, L. and {Raison}, F. and {Rebolo}, R. and {Renzi}, A. and {Rhodes}, J. and {Riccio}, G. and {Romelli}, E. and {Roncarelli}, M. and {Rossetti}, E. and {Saglia}, R. and {Sakr}, Z. and {S{\'a}nchez}, A.~G. and {Sapone}, D. and {Schneider}, P. and {Schrabback}, T. and {Secroun}, A. and {Seidel}, G. and {Serrano}, S. and {Sirignano}, C. and {Sirri}, G. and {Skottfelt}, J. and {Stanco}, L. and {Steinwagner}, J. and {Tallada-Cresp{\'\i}}, P. and {Tavagnacco}, D. and {Taylor}, A.~N. and {Tereno}, I. and {Toledo-Moreo}, R. and {Torradeflot}, F. and {Tutusaus}, I. and {Valentijn}, E.~A. and {Valenziano}, L.},
        title = "{Euclid: The Early Release Observations Lens Search Experiment}",
      journal = {\aap},
         year = 2025,
        month = may,
       volume = {697},
          eid = {A14},
        pages = {A14},
          doi = {10.1051/0004-6361/202451868},
archivePrefix = {arXiv},
       eprint = {2408.06217},
 primaryClass = {astro-ph.GA},
       adsurl = {https://ui.adsabs.harvard.edu/abs/2025A&A...697A..14A}
}

@article{vonHausegger:2024jan,
    author = "von Hausegger, Sebastian",
    title = "{The expected kinematic matter dipole is robust against source evolution}",
    eprint = "2404.07929",
    archivePrefix = "arXiv",
    primaryClass = "astro-ph.CO",
    doi = "10.1093/mnrasl/slae092",
    journal = "Mon. Not. Roy. Astron. Soc.",
    volume = "535",
    number = "1",
    pages = "L49--L53",
    year = "2024"
}

@article{Lacasa:2024ybp,
    author = "Lacasa, Fabien and Bonvin, Camille and Dalang, Charles and Durrer, Ruth",
    title = "{Fast and spurious: a robust determination of our peculiar velocity with future galaxy surveys}",
    eprint = "2402.18438",
    archivePrefix = "arXiv",
    primaryClass = "astro-ph.CO",
    doi = "10.1088/1475-7516/2024/06/045",
    journal = "JCAP",
    volume = "06",
    pages = "045",
    year = "2024"
}

@article{vonHausegger:2024fcu,
    author = "von Hausegger, Sebastian and Dalang, Charles",
    title = "{Redshift tomography of the kinematic matter dipole}",
    eprint = "2412.13162",
    archivePrefix = "arXiv",
    primaryClass = "astro-ph.CO",
    doi = "10.1103/PhysRevD.111.123547",
    journal = "Phys. Rev. D",
    volume = "111",
    number = "12",
    pages = "123547",
    year = "2025"
}

@ARTICLE{Knabel2025,
       author = {{Knabel}, Shawn and {Mozumdar}, Pritom and {Shajib}, Anowar J. and {Treu}, Tommaso and {Cappellari}, Michele and {Spiniello}, Chiara and {Birrer}, Simon},
        title = "{TDCOSMO: XIX. Measuring stellar velocity dispersion with sub-percent accuracy for cosmography}",
      journal = {\aap},
         year = 2025,
        month = nov,
       volume = {703},
          eid = {A117},
        pages = {A117},
          doi = {10.1051/0004-6361/202554229},
archivePrefix = {arXiv},
       eprint = {2502.16034},
 primaryClass = {astro-ph.GA},
       adsurl = {https://ui.adsabs.harvard.edu/abs/2025A&A...703A.117K}
}

@ARTICLE{TDCOSMO2025,
       author = {{Tdcosmo Collaboration} and {Birrer}, Simon and {Buckley-Geer}, Elizabeth J. and {Cappellari}, Michele and {Courbin}, Fr{\'e}d{\'e}ric and {Dux}, Fr{\'e}d{\'e}ric and {Fassnacht}, Christopher D. and {Frieman}, Joshua A. and {Galan}, Aymeric and {Gilman}, Daniel and {Huang}, Xiang-Yu and {Knabel}, Shawn and {Langeroodi}, Danial and {Lin}, Huan and {Millon}, Martin and {Morishita}, Takahiro and {Motta}, Veronica and {Mozumdar}, Pritom and {Paic}, Eric and {Shajib}, Anowar J. and {Sheu}, William and {Sluse}, Dominique and {Sonnenfeld}, Alessandro and {Spiniello}, Chiara and {Stiavelli}, Massimo and {Suyu}, Sherry H. and {Tan}, Chin Yi and {Treu}, Tommaso and {van de Vyvere}, Lyne and {Wang}, Han and {Wells}, Patrick and {Williams}, Devon M. and {Wong}, Kenneth C.},
        title = "{TDCOSMO 2025: Cosmological constraints from strong lensing time delays}",
    collaboration = "TDCOSMO",
      journal = {\aap},
         year = 2025,
        month = dec,
       volume = {704},
          eid = {A63},
        pages = {A63},
          doi = {10.1051/0004-6361/202555801},
archivePrefix = {arXiv},
       eprint = {2506.03023},
 primaryClass = {astro-ph.CO},
       adsurl = {https://ui.adsabs.harvard.edu/abs/2025A&A...704A..63T}
}

@ARTICLE{Oayda2025,
       author = {{Oayda}, Oliver T. and {Mittal}, Vasudev and {Lewis}, Geraint F.},
        title = "{Cosmic multipoles in galaxy surveys - I. How inferences depend on source counts and masks}",
      journal = {\mnras},
         year = 2025,
        month = feb,
       volume = {537},
       number = {1},
        pages = {1-20},
          doi = {10.1093/mnras/stae2776},
archivePrefix = {arXiv},
       eprint = {2412.12600},
 primaryClass = {astro-ph.CO},
       adsurl = {https://ui.adsabs.harvard.edu/abs/2025MNRAS.537....1O}
}

@ARTICLE{Land-Strykowski2025,
       author = {{Land-Strykowski}, Mali and {Lewis}, Geraint F. and {Murphy}, Tara},
        title = "{Cosmic dipole tensions: confronting the cosmic microwave background with infrared and radio populations of cosmological sources}",
      journal = {\mnras},
         year = 2025,
        month = nov,
       volume = {543},
       number = {4},
        pages = {3229-3241},
          doi = {10.1093/mnras/staf1621},
archivePrefix = {arXiv},
       eprint = {2509.18689},
 primaryClass = {astro-ph.CO},
       adsurl = {https://ui.adsabs.harvard.edu/abs/2025MNRAS.543.3229L}
}

@ARTICLE{Krywonos2024,
       author = {{Krywonos}, Jordan and {Hotinli}, Selim C. and {Johnson}, Matthew C.},
        title = "{Constraints on cosmology beyond $\Lambda$CDM with kinetic Sunyaev Zel'dovich velocity reconstruction}",
      journal = {arXiv e-prints},
         year = 2024,
        month = aug,
          eid = {arXiv:2408.05264},
        pages = {arXiv:2408.05264},
          doi = {10.48550/arXiv.2408.05264},
archivePrefix = {arXiv},
       eprint = {2408.05264},
 primaryClass = {astro-ph.CO},
       adsurl = {https://ui.adsabs.harvard.edu/abs/2024arXiv240805264K}
}

@ARTICLE{Bloch2024,
       author = {{Bloch}, Richard and {Johnson}, Matthew C.},
        title = "{Kinetic Sunyaev Zel'dovich velocity reconstruction from Planck and unWISE}",
      journal = {arXiv e-prints},
         year = 2024,
        month = may,
          eid = {arXiv:2405.00809},
        pages = {arXiv:2405.00809},
          doi = {10.48550/arXiv.2405.00809},
archivePrefix = {arXiv},
       eprint = {2405.00809},
 primaryClass = {astro-ph.CO},
       adsurl = {https://ui.adsabs.harvard.edu/abs/2024arXiv240500809B}
}

@ARTICLE{Bolton2008,
       author = {{Bolton}, Adam S. and {Treu}, Tommaso and {Koopmans}, L{\'e}on V.~E. and {Gavazzi}, Rapha{\"e}l and {Moustakas}, Leonidas A. and {Burles}, Scott and {Schlegel}, David J. and {Wayth}, Randall},
        title = "{The Sloan Lens ACS Survey. VII. Elliptical Galaxy Scaling Laws from Direct Observational Mass Measurements}",
      journal = {\apj},
         year = 2008,
        month = sep,
       volume = {684},
       number = {1},
        pages = {248-259},
          doi = {10.1086/589989},
archivePrefix = {arXiv},
       eprint = {0805.1932},
 primaryClass = {astro-ph},
       adsurl = {https://ui.adsabs.harvard.edu/abs/2008ApJ...684..248B}
}

@ARTICLE{Bezanson2013,
       author = {{Bezanson}, Rachel and {van Dokkum}, Pieter G. and {van de Sande}, Jesse and {Franx}, Marijn and {Leja}, Joel and {Kriek}, Mariska},
        title = "{Tight Correlations between Massive Galaxy Structural Properties and Dynamics: The Mass Fundamental Plane was in Place by z \raisebox{-0.5ex}\textasciitilde 2}",
      journal = {\apjl},
         year = 2013,
        month = dec,
       volume = {779},
       number = {2},
          eid = {L21},
        pages = {L21},
          doi = {10.1088/2041-8205/779/2/L21},
archivePrefix = {arXiv},
       eprint = {1309.6638},
 primaryClass = {astro-ph.CO},
       adsurl = {https://ui.adsabs.harvard.edu/abs/2013ApJ...779L..21B}
}
